\documentclass[a4paper,nofootinbib,reprint,nobibnotes,superscriptaddress,enumitem]{revtex4-1} 

\usepackage[english]{babel}
 \addto\captionsenglish{
                       }

\usepackage{graphicx}
 \graphicspath{{graphics/}}

\usepackage[caption=false]{subfig}
\usepackage[export]{adjustbox}

\usepackage{amsmath}
\usepackage{amssymb}
\usepackage[version=3]{mhchem}
\usepackage{mathrsfs}
\usepackage{bm}

\usepackage{upgreek}

\usepackage[hidelinks,linktocpage]{hyperref}
\hypersetup{
            colorlinks,
            linkcolor={red!70!black},
            citecolor={green!70!black},
            urlcolor={blue!70!black}
           }

\usepackage[usenames,dvipsnames]{xcolor}
\usepackage{comment} 

\usepackage{xspace}

\newcommand{\al}  {\ensuremath{\alpha}\xspace}
\newcommand{\bt}  {\ensuremath{\beta}\xspace}
\newcommand{\gm}  {\ensuremath{\gamma}\xspace}

\newcommand{\bb}  {\ensuremath{0\nu\beta\beta}\xspace}
\newcommand{\bbvv}{\ensuremath{2\nu\beta\beta}\xspace}

\newcommand{\MI}  {\ensuremath{\mathcal M 1}\xspace}
\newcommand{\MII} {\ensuremath{\mathcal M 2}\xspace}
\newcommand{\MIII}{\ensuremath{\mathcal M 3}\xspace}

\newcommand{\QF}  {QF\xspace}

\newcommand{\MC}      {Monte Carlo\xspace}
\newcommand{\ares}    {{\tt Ares}\xspace}
\newcommand{\JAGS}    {JAGS\xspace}
\newcommand{\qshields}{{\tt QShields}\xspace}

\newcommand{\CUORE} {CUORE\xspace}
\newcommand{\CUOREZ}{CUORE-0\xspace}
\newcommand{\CUPID} {CUPID\xspace}

\newcommand{\ELS}   {ELS\xspace}

\newcommand{\HEX}   {HEX\xspace}
\newcommand{\ILS}   {ILS\xspace}
\newcommand{\MX}    {MC\xspace}
\newcommand{\TL}    {TL\xspace}
\newcommand{\TSP}   {TSP\xspace}

\newcommand{\xtals}   {Crystals\xspace}
\newcommand{\CuNOSV}  {Close parts\xspace}
\newcommand{\CuOFEIn} {Inner shields\xspace}
\newcommand{\CuOFEOut}{Outer shields\xspace}
\newcommand{\RomanPb} {ILS\xspace}
\newcommand{\ModPbIn} {TL\xspace}
\newcommand{\ModPbOut}{ELS\xspace}

\begin{document}

\title{Data-driven background model for the CUORE experiment}
\author{D.~Q.~Adams}
\affiliation{Department of Physics and Astronomy, University of South Carolina, Columbia, SC 29208, USA}

\author{C.~Alduino}
\affiliation{Department of Physics and Astronomy, University of South Carolina, Columbia, SC 29208, USA}

\author{K.~Alfonso}
\affiliation{Center for Neutrino Physics, Virginia Polytechnic Institute and State University, Blacksburg, Virginia 24061, USA}

\author{F.~T.~Avignone~III}
\affiliation{Department of Physics and Astronomy, University of South Carolina, Columbia, SC 29208, USA}

\author{O.~Azzolini}
\affiliation{INFN -- Laboratori Nazionali di Legnaro, Legnaro (Padova) I-35020, Italy}

\author{G.~Bari}
\affiliation{INFN -- Sezione di Bologna, Bologna I-40127, Italy}

\author{F.~Bellini}
\affiliation{Dipartimento di Fisica, Sapienza Universit\`{a} di Roma, Roma I-00185, Italy}
\affiliation{INFN -- Sezione di Roma, Roma I-00185, Italy}

\author{G.~Benato}
\affiliation{Gran Sasso Science Institute, L'Aquila I-67100, Italy}
\affiliation{INFN -- Laboratori Nazionali del Gran Sasso, Assergi (L'Aquila) I-67100, Italy}

\author{M.~Beretta}
\affiliation{Department of Physics, University of California, Berkeley, CA 94720, USA}

\author{M.~Biassoni}
\affiliation{INFN -- Sezione di Milano Bicocca, Milano I-20126, Italy}

% ...
\date{\today}

\begin{abstract}
 We present the model we developed to reconstruct the \CUORE radioactive background based on the analysis of an experimental exposure of $1038.4$ kg yr. The data reconstruction relies on a simultaneous Bayesian fit applied to energy spectra over a broad energy range.  The high granularity of the \CUORE detector, together with the large exposure and extended stable operations, allow for an in-depth exploration of both spatial and time dependence of backgrounds.
 We achieve high sensitivity to both bulk and surface activities of the materials of the setup, detecting levels as low as 10 nBq kg$^{-1}$ and 0.1 nBq cm$^{-2}$, respectively.
 We compare the contamination levels we extract from the background model with prior radio-assay data, which informs future background risk mitigation strategies. The results of this background model play a crucial role in constructing the background budget for the \CUPID experiment as it will exploit the same CUORE infrastructure.
\end{abstract}

\maketitle

 \section{Introduction}
 \label{sec:intro}
 
The Cryogenic Underground Observatory for Rare Events (\CUORE~\cite{CUORE:2021ctv}) is a tonne-scale cryogenic detector located at Laboratori Nazionali del Gran Sasso (Italy). The primary physics goal of the experiment is to search for neutrinoless double-beta decay (\bb) of \ce{^{130}Te}~\cite{Agostini:2022zub}.
\bb is a hypothetical lepton-number-violating process that, if observed, would demonstrate that neutrinos are Majorana fermions and that lepton number is not a symmetry of nature.
Due to the low-background and excellent energy resolution, the experiment is well-placed to study other rare decays such as the Standard-Model-allowed two-neutrino double-beta decay (\bbvv) of \ce{^{130}Te} and to search for exotic phenomena such as CPT or Lorentz non-conservation, charge-violating phenomena, and \bb via Majoron emission~\cite{CUORE:2018tky}. 

A thorough understanding of the \CUORE background is essential to perform these studies. For example, a robust reconstruction of the background components allows a precise measurement of both half-life and spectral shape of \bbvv~of \ce{^{130}Te}~\cite{CUORE:2020bok}. This provides a benchmark to validate approximation methods employed to calculate double-$\beta$ nuclear matrix elements. In addition, much of the CUORE cryogenic infrastructure will be used to host a next-generation experiment, the \CUORE upgrade with Particle IDentification (\CUPID~\cite{CUPID_bsl_paper}). Thus, the characterization of the background by this system on the \CUORE detector is essential to build a robust, data-driven background budget for the \CUPID experiment. 
 
In this work, we present a comprehensive description of the model used to reconstruct the \CUORE data and provide a detailed assessment of the radioactive contamination of the cryogenic infrastructure and detector components based on a Bayesian analysis, which includes prior information from materials screening carried out as part of CUORE construction.  
\section{Experiment Overview}
\label{sec:experiment_overview}
In this section we briefly summarize the CUORE detector and cryogenic infrastructure, as well as the data selection and production, which are needed for context to describe the model we developed to reconstruct the data. More complete and detailed descriptions of the experiment and infrastructure are available in~\cite{Alduino:2016vjd,Alessandria:2013ufa,Alduino:2019xia,DAddabbo:2017efe}. 
 
\begin{figure*}[t]
  \centering
  \includegraphics[width=2.\columnwidth]{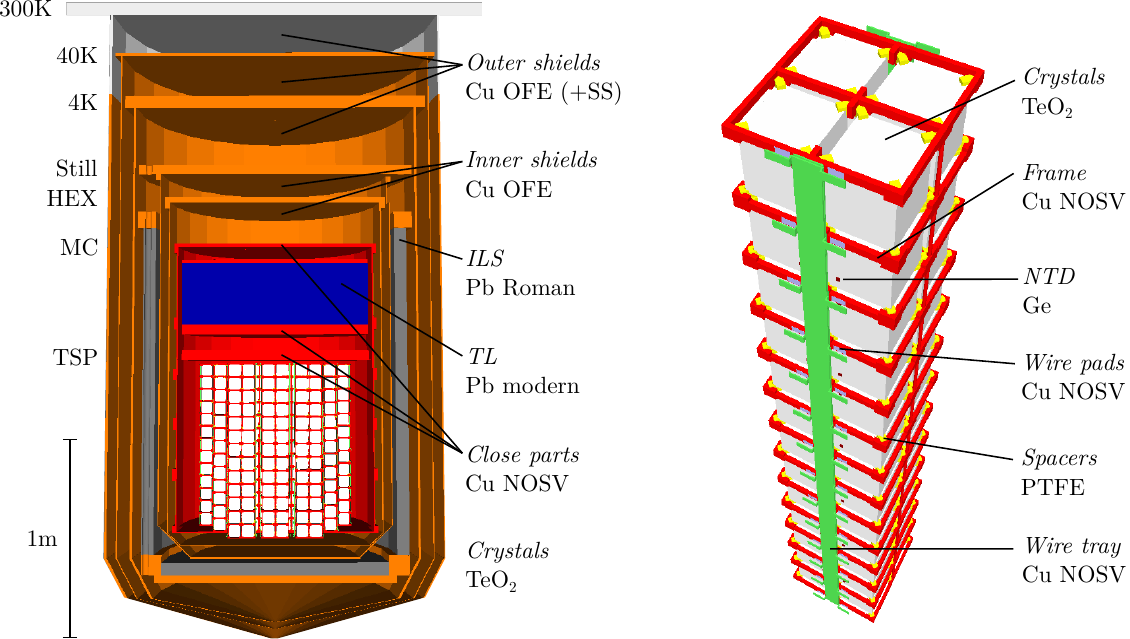}
  \caption{(\textit{Left}) Rendering of the \CUORE cryostat and detector as implemented in the \MC simulation; the actual length scale is reported as a reference. On the left, the different thermal stages are reported; on the right, the volumes are presented as grouped in the \MC, together with the material they are made of. The \ce{TeO_{2}} crystals are depicted in white, the NOSV copper components in red, the OFE copper in orange, the \TL in blue, the \RomanPb in gray and the 300K stainless steel in light gray.
  (\textit{Right}) Detailed view of a \CUORE tower, where all the different components are showed in different colors: \ce{TeO_{2}} crystals in white, NOSV-copper frames in red, PTFE supports in yellow, NOSV-copper wire trays in green and NOSV-copper wire pads in blue.}
  \label{fig:cryostat}
 \end{figure*}

 The \CUORE detector (Fig.~\ref{fig:cryostat}) is a close-packed array of $988$ \ce{TeO_2} individual crystals operated as cryogenic calorimeters (also called bolometers) arranged into $19$ towers of $13$ four-crystal floors, i.\,e.\ $52$ crystals per tower. Each bolometer is instrumented with a \ce{Ge} Neutron Transmutation Doped (NTD) thermistor to measure the temperature, and with a \ce{Si} heater to stabilize the detector gain against long-term temperature drifts induced by the cryogenic system. The crystals are supported in the tower structure by a set of copper frames and held in position by Polytetrafluoroethylene (PTFE) spacers. The path for the electrical connection between the on-crystal instrumentation and the front-end electronics~\cite{Arnaboldi:2015wvc,Arnaboldi:2017aek} is provided by copper traces deposited on flexible PEN substrates which span the length of each tower.  The detector array is anchored to a copper Tower Support Plate (\TSP) placed at the center of a custom \ce{^3He}/\ce{^4He} dilution refrigerator that allows the operation of the bolometers at a temperature of about $10$ mK.
 
 The \CUORE cryostat comprises six nested copper vessels, which thermalize at decreasing temperatures from room temperature down to $10$ mK.
 Each thermal stage is named for its approximate temperature or by the corresponding component of the dilution unit: $300$ K, $40$ K, $4$ K, $800$ mK or Still, $50$ mK or Heat EXchanger (\HEX)
 and $10$ mK or Mixing Chamber (\MX).
 Inside the cryostat, two lead shields protect the detector from the external radioactivity: the Inner Lead Shield (\ILS) is suspended between the $4$ K and the Still stages and provides shielding both laterally as well as from below. The Top Lead (\TL) is positioned below the \MX plate and provides shielding from the cryogenic apparatus above.
 Outside the cryostat, the room temperature External Lead Shield (\ELS) and a neutron shield (made of polyethylene and a layer of boric acid) provide additional shielding from the side and from below.

To minimize background in the experiment, radio-pure materials were selected through dedicated assay campaigns~\cite{Alduino:2017qet}. \emph{Ultra}-cleaning treatments were developed and applied to the corresponding material~\cite{Alessandria:2012zp} to mitigate background induced from residual \al decays on critical surfaces. Furthermore, storage and handling protocols were implemented to minimize recontamination during assembly, installation and commissioning of the detector array~\cite{Buccheri:2014bma,Benato:2017kdf}.

\CUORE began taking data in April 2017 and, to date, more than $2$ t\,yr of \ce{TeO_2} exposure have been collected~\cite{CUORE_talk_taup}.  The data collection is organized into \emph{datasets}, which we define as accumulations of about $1-2$ months of so-called physics runs sandwiched between a few days of calibration runs. The physics runs are used for the \bb search and other studies, including modeling the background sources.
The set of data considered for this work corresponds to about a half of the collected exposure, specifically $1038.4$ kg yr, and is the same exposure previously analyzed to search for \bb decay in 2022~\cite{CUORE:2021mvw}.

For each detector, we acquire and save a continuous data stream.
We trigger thermal pulses by means of the optimum trigger, a trigger algorithm based on the optimum filter (OF~\cite{Gatti:1986cw}), that allows to maximize the signal-to-noise ratio in the frequency domain. We then define 10 s time windows (which include 3 s pre-trigger) around the triggered pulse and we apply the OF to compute the pulse amplitude.
Moreover, we correct for gain fluctuations caused by temperature drifts of the system by continually monitoring the detector response to mono-energetic heater pulses~\cite{Alduino:2016zrl}.
The stabilized pulse amplitudes are converted to energy values by using dedicated calibration data.
Subsequently, we apply a series of event-selection criteria in order to exclude non-physical events. These criteria include temporal cuts to eliminate periods of hardware malfunctions, pulse quality cuts and pulse-shape cuts based on a principal component analysis (PCA). 
Leveraging on the compact design of our detector, where adjacent-crystal distances span between $\sim$ 8 mm and $\sim$ 5 cm (for crystals on different towers), we build multiplets of multi-crystal events occurring within a $\mathcal{O}$(ms) time window.
Depending on the specific analysis, we can use the different multiplets in order to exploit several event topologies that likely share a common physics source.
 \section{Background sources and \MC simulations}
 \label{sec:MC}
 
 The CUORE background events passing the data-quality selections originate from radioactive contaminants in the experimental setup or from particle fluxes in the external environment, that we call background sources. Each background source produces energy spectra with distinctive features, such as peaks due to \gm-rays and \al-particle interactions, continuous spectra from \bt decays or structures due to multi-site events or emissions in time coincidence, depending on their location and strength.
 Moreover, thanks to the modularity of our detector, we exploit signals detected in time coincidence to assess the contamination producing \al-decays on crystal surfaces. This also allows maximizing the information on background sources which produce multi-crystal events.
 The \CUORE background model aims to determine the activity of the different background sources, by disentangling their contribution to the experimental spectra. This is done by fitting to the data a linear combination of the background sources' induced spectra, obtained with \MC simulations. The production of the \MC is a two-step process.
 First, the simulations for each hypothetical source within the setup are generated with \qshields, a Geant4~\cite{GEANT4:2002zbu,Allison:2006ve} application that simulates particle propagation and interaction throughout the \CUORE cryostat and detector; we make use of the standard physics lists
 QGSP\_BERT\_HP and Livermore\_EM, for hadronic and electromagneic processes, respectively. 
 Then, the outputs are processed by a software tool, named \ares, which applies the detector response to the raw \MC and provides as output simulated events which resemble real data acquired with \CUORE.

 \subsection{Simulation production}
 \label{sec:qshields}
 
A realistic description of the \CUORE geometry and materials in the simulation is crucial for the construction of the background model.
In \qshields, all the elements described in Section~\ref{sec:experiment_overview} are implemented in the geometrical description. 
We run a \MC simulation for each background source identified by a preliminary analysis of the data.
In each simulation, a specific radionuclide (or decay chain) is generated in one of the volumes of the geometry, whereas the particle propagation is always considered for the whole geometry. 
The same contaminant in different volumes can result in very degenerate background spectra, whose differences are below the statistical uncertainty of the experimental data and cannot be disentangled.
Therefore, some adjacent volumes are grouped together as having the same contamination if they give completely degenerate spectra. Moreover, setup components which are made of the same grade of material and underwent the same cleaning treatments are assumed to have equal bulk and surface contamination.

We designate the following seven main source volumes or volume groups:

 \begin{enumerate}
 \item {\it\xtals}: the \ce{TeO_{2}} crystals, excluding the NTDs and heaters as they contribute negligibly to the background~\cite{Alduino:2017qet}.
 \item {\it\CuNOSV}: refer to parts which are close to the detectors, with and without direct line-of-sight to the crystals. Items with direct line-of-sight include the PTFE spacers holding the crystals and copper parts (tower frames, supports for readout wires, tiles covering the inside of the \MX shield, guide-tubes of the calibration system \cite{Cushman:2016cnv}).  Items with no direct line-of-sight are the \MX shield, the \TSP, and the plates that sandwich the \TL. The copper in the {\it\CuNOSV} volume is electrolytic tough-pitch copper, known commercially as NOSV copper~\cite{CuNOSV_aurubis}, and was selected for its low radioactivity and high thermal conductivity at low temperatures, crucial for the operation of the detector. Although the \MX shield is also made of NOSV copper, it underwent a different surface cleaning procedure. Therefore, we assign it the same bulk contamination activity as the other close components but treat its surface contamination separately. 
 PTFE spacers and other NOSV copper parts underwent different ultra-cleaning~\cite{Alduino:2016vjd}. Nonetheless, the total mass of PTFE is significantly lower than that of close NOSV-copper tower components and dedicated study could not distinguish their background contribution; therefore PTFE contaminants have been neglected.
 \item {\it\CuOFEIn}: thermal radiation shields between the \MX and the \ILS, namely the \HEX and Still vessels and respective top plates. They are made from oxygen-free electrolytic (OFE) C10100 copper. 
 \item {\it\RomanPb}: shield made of ancient Roman lead which is depleted of \ce{^{210}Pb}~\cite{Alessandrello:1998RL}.
 \item {\it\CuOFEOut:} thermal radiation shields outside the \ILS, namely the $4$ K, $40$ K, and $300$ K stage vessels and their respective top plates made of OFE copper. This source volume also includes the stainless-steel cap of the 300 K vessel and superinsulation installed in the cryostat as their contributions to the background are completely degenerate.
 \item {\it\ModPbIn}: shield made of specially selected low-radioactivity lead.
 \item {\it\ModPbOut}: shield made of commercial low-radioactivity lead but different from that of \ModPbIn.
\end{enumerate}
 
 \qshields can generate and propagate photons, electrons, positrons, \al particles, nuclear recoils, neutrons and muons. All primary particles and their resulting secondaries are propagated down to keV energies. Nuclear transitions are based on a customized implementation of the {\it G4RadioactiveDecay} database, which keeps track of the time correlations in the radioactive chains.
 Bulk contamination is simulated assuming the radio-nuclides uniformly distributed inside the volumes. The only exception to this is our treatment of \ce{^{40}K} contamination in one of the towers of the {\it\xtals} volume, which is further described in App.~\ref{app:K_t12}.
 Surface contamination is assumed uniform over the surfaces and the corresponding radioactivity concentrations follow exponentially-decreasing depth ($d$) profiles $\textrm{exp}(-\lambda \, d)$, where $\lambda$ is the characteristic depth. To better model the shape of structures we observe in the data associated with \al decays, we consider depths spanning from the nm to the tens-of-$\mu$m scale.

 The identification of the contaminants to be simulated is the combined result of an extensive campaign of radio-assay measurements on different materials. Furthermore, the expectations from the modeling of the \CUOREZ background~\cite{Alduino:2016vtd}, and the search for distinctive features in the \CUORE experimental data contributed to this identification.
 
 The full list of contaminants in each volume is reported in Table~\ref{tab:contaminants}.
 The complete decay chains of \ce{^{238}U} and \ce{^{232}Th} are simulated for all volumes, except for {\it \ModPbOut}.
 For both {\it \ModPbIn} and {\it \ModPbOut}, the lower part of the \ce{^{238}U} decay chain, from \ce{^{210}Bi} to \ce{^{206}Pb} is simulated, assuming it in equilibrium with a contamination of \ce{^{210}Pb}.%
 \footnote{In order to save computational time, we neglect the decay of the \ce{^{210}Pb} itself, since it induces a negligible background.}
 The decay chain of \ce{^{235}U} is simulated only for {\it\xtals} and {\it\CuNOSV} because of its very weak signatures in the data; its activity is fixed with respect to that of \ce{^{238}U} by their natural activity ratio.
 In \ce{TeO_{2}} and copper we simulate the primordial radio-nuclide \ce{^{40}K}.
 We consider cosmogenic-activation isotopes, namely \ce{^{125}Sb}, \ce{^{110m}Ag}, \ce{^{108m}Ag} and \ce{^{60}Co} in \ce{TeO_{2}}, and \ce{^{60}Co} and \ce{^{54}Mn} in copper.
 We also include the fallout products \ce{^{207}Bi} in lead (found during radio-assays) and \ce{^{137}Cs} in copper (observed in the past \CUOREZ data).
 Other contaminants of \ce{TeO_{2}} are the long-lived \ce{^{190}Pt}, which is contained in macroscopic residues inside the crystals as left-over of the platinum crucibles used during the growth process~\cite{Arnaboldi:2010fj}, and a possible contamination of \ce{^{147}Sm} (see App.~\ref{app:TeO2OtherContaminants}).
 Lastly, we include in {\it\xtals} the decay of \ce{^{130}Te} via \bbvv assuming the single-state-dominance model to simulate its spectrum.
 This work is not devoted to the precise determination of the \bbvv spectral shape and this choice is not impacting the results; previous studies have hinted to a preference for this model~\cite{CUORE:2020bok}.

 Backgrounds outside the experimental apparatus (neutrons and environmental \gm-rays) are shielded at a level that makes them completely negligible for reconstructing the data. 
 The only simulation that does not originate from a volume in the CUORE geometry is the cosmic muon flux.
 
 \begin{center}
  \begin{table*}[t]
   \caption{List of the simulated volumes and contaminants. The arrows indicate fractions of a decay chain assumed to be in secular equilibrium. The partial chain \ce{^{210}Bi \to ^{206}Pb} has been used to simulate the contamination in equilibrium with \ce{^{210}Pb} in {\it\ILS} and {\it\ModPbOut}. This has been done to save computational time since the contribution from the decay of \ce{^{210}Pb} alone is negligible for these volumes. The {\it\MX} volume has been isolated from {\it\CuNOSV} in dealing with superficial contamination because the \MX thermal shield underwent a different surface cleaning due to its large size. The internal surface of the shield is covered with tiles, hence the background induced by the \MX surface contamination mainly comes from the \bt/\gm radiation of the decay chains.
   Analogously, we simulate a surface contamination of \ce{^{210}Pb} on the {\it\HEX} volume.}
     \begin{tabular}{l l l}
      Volume                  &Material        &\quad Contaminants \\
      \hline  \\[-8pt] {\it Bulk} \\ \cline{1-1} \\[-8pt]
      \xtals                  &\ce{TeO_2}      &\quad\ce{^{130}Te} \bbvv, \ce{^{232}Th} / \ce{^{228}Ra \to ^{208}Pb}, \ce{^{238}U \to ^{230}Th} / \ce{^{230}Th} / \ce{^{226}Ra \to ^{210}Pb} / \\
                              &                &\quad\ce{^{210}Pb \to ^{206}Pb}, \ce{^{235}U \to ^{231}Pa} / \ce{^{231}Pa \to ^{207}Pb}, \ce{^{190}Pt}, \ce{^{147}Sm}, \ce{^{125}Sb}, \\
                              &                &\quad\ce{^{110m}Ag}, \ce{^{108m}Ag}, \ce{^{60}Co}, \ce{^{40}K} \\[10pt]
      \CuNOSV                 &\ce{Cu} NOSV    &\quad\ce{^{232}Th \to ^{208}Pb}, \ce{^{238}U \to ^{206}Pb}, \ce{^{235}U \to ^{207}Pb}, \ce{^{137}Cs}, \ce{^{60}Co}, \ce{^{54}Mn}, \ce{^{40}K}                    \\[10pt]
      \CuOFEIn                &\ce{Cu} OFE     &\quad\ce{^{232}Th \to ^{208}Pb}, \ce{^{238}U \to ^{206}Pb}, \ce{^{137}Cs}, \ce{^{60}Co}, \ce{^{54}Mn}, \ce{^{40}K}                    \\[10pt]
      \RomanPb                &\ce{Pb} Roman   &\quad\ce{^{232}Th \to ^{208}Pb}, \ce{^{238}U \to ^{206}Pb}, \ce{^{108m}Ag}                                                            \\[10pt]
      \CuOFEOut               &\ce{Cu} OFE     &\quad\ce{^{232}Th \to ^{208}Pb}, \ce{^{238}U \to ^{206}Pb}, \ce{^{137}Cs}, \ce{^{60}Co}, \ce{^{54}Mn}, \ce{^{40}K}                    \\
                              &Stainless steel & \\
                              &Superinsulation & \\[10pt]

      \ModPbIn                &\ce{Pb}         &\quad\ce{^{232}Th \to ^{208}Pb}, \ce{^{238}U \to ^{206}Pb}, \ce{^{210}Bi \to ^{206}Pb}                                                \\[10pt]
      \ModPbOut               &\ce{Pb}         &\quad\ce{^{210}Bi \to ^{206}Pb}, \ce{^{207}Bi}                                                                                        \\[10pt]
      (external)              &--              &\quad cosmic-$\mu$ flux                                                                                                               \\[5pt]
      {\it Surface} \\ \cline{1-1} \\[-8pt]
      \xtals                  &\ce{TeO_2}      &\quad\ce{^{232}Th} / \ce{^{228}Ra \to ^{208}Pb}, \ce{^{238}U \to ^{230}Th} / \ce{^{230}Th} / \ce{^{226}Ra \to ^{210}Pb} / \\
                              &                &\quad\ce{^{210}Pb  \to ^{206}Pb}, \ce{^{235}U \to ^{231}Pa} / \ce{^{231}Pa \to ^{207}Pb}                                              \\[10pt]
      \CuNOSV (no \MX)~       &\ce{Cu} NOSV    &\quad\ce{^{232}Th \to ^{208}Pb}, \ce{^{238}U \to ^{206}Pb}, \ce{^{210}Pb  \to ^{206}Pb}, \ce{^{235}U \to ^{207}Pb} \\[10pt]
      \MX                     &\ce{Cu} NOSV    &\quad\ce{^{232}Th \to ^{208}Pb}, \ce{^{238}U \to ^{206}Pb} \ce{^{210}Pb  \to ^{206}Pb}, \ce{^{235}U \to ^{207}Pb}                                               \\[10pt]
      \HEX                    &\ce{Cu} OFE     &\quad\ce{^{210}Pb  \to ^{206}Pb}                                                                                                      \\[5pt]
      \hline
     \end{tabular}
    \label{tab:contaminants}
   \end{table*}
  \end{center}
  
 \subsection{Simulation processing}
 \label{sec:ares}
 
 The \qshields output undergoes a post-processing phase implemented in \ares, to convert the raw \MC into \CUORE-like data, accounting for detector response effects and data selection cuts made in the analysis~\cite{Alduino:2016zrl,CUORE_analysis}.
 In particular, we assign a dataset and a timestamp to the events and include dataset-dependent information, i.\,e.\ detector energy resolution, the status of the individual channels (for example active or inactive) and the event-selection efficiencies.
 In the case of \al particles, we apply a quenching factor (\QF) as described in App.~\ref{app:TeO2OtherContaminants}.

 We account for unresolvable pile-up effects by merging energy depositions occurring in the same crystal within a time window of $5$ ms; the outcome is a single event with energy equal to the sum of the individual depositions associated with the event.
 We discard the resolvable pile-up, i.\,e.\ events occurring in the same crystal with a time distance larger than $5$ ms but with overlapping acquisition windows.
 We select events with reconstructed energies greater than $40$ keV. 
 This threshold is sufficiently low to fully include structures produced by recoiling nuclei in low-Q-value \al decays, while still high enough to exclude the energy region where our control of efficiencies is limited. We also discard events with reconstructed energies larger than $10$~MeV, at the limit of our detector dynamic range (saturated events).
 
 Finally, we compute the event {\it multiplicity} defined as follows: events isolated in time and space are labeled as multiplicity~$1$ (\MI); events occurring inside a time window of \mbox{$\pm 30$ ms} and involving neighboring crystals closer than $15$ cm are grouped into higher multiplicities labeled \MII, \MIII and so on according to the number of crystals involved. The introduction of the multiplicity label is based on the assumption that events close in space and time likely originate from the same physical process. The sum of the individual energies in the same multiplet is referred to as total-energy of the event.
 
 The simulated spectra at different multiplicities are then used to reconstruct the corresponding ones built with the \CUORE data.
 \section{Fit procedure}
 \label{sec:fit}
 
 The CUORE data used to build the background model are organized into a collection of binned energy histograms containing \MI and \MII events.
 We currently do not consider higher-multiplicity spectra, as we find they provide little additional information to the background model. The only exception is given by the high-multiplicity data used to fix the prior distribution of the muon-induced background (see Sec.~\ref{sec:priors}).  
 
 We assume that the number of counts in each bin follows a Poisson probability distribution $Pois(n,\nu)$, where $n$ is the number of events observed by \CUORE and $\nu$ is the expected value. The latter is defined as a linear combination of the bin counts in the simulated spectra coming from the different background sources, each weighted by its normalization factor.  
 Therefore, considering the $i$-th bin for the energy spectrum $\kappa$ we can write:
 \begin{equation}
     \nu_{\kappa,i} = \sum_{j}N_{j}(w_{\kappa,i})_{j},
 \end{equation}
where index $\kappa$ runs over the collection of histograms into which the data are organized, including both \MI and \MII events, $j$ runs over the background sources, $N_j$ is the normalization factor of source $j$, and $(w_{\kappa,i})_j$ is the $i$-th bin content of the spectrum $\kappa$ for the source $j$. 
The total likelihood takes the form
 \begin{equation}
    \label{eq:TotalLikelihood}
     \mathcal{L}(\{N_{j}\} \,\,|\,\, \text{data}) = \prod_{\kappa}\prod_{i} Pois(n_{\kappa,i},\nu_{\kappa,i})
 \end{equation}
 and the normalization factors are the fit parameters.
 We assign a prior probability distribution to each $N_{j}$ (see Sec.~\ref{sec:priors}) and through Bayesian statistical inference using the likelihood outlined in Eq.~\eqref{eq:TotalLikelihood}, we sample the multidimensional posterior. 
 The sampling procedure is managed by a Markov-Chain \MC through a Gibbs sampling algorithm implemented in the \JAGS software~\cite{JAGS}.%
 \footnote{The JAGS-based analysis tool has been firstly developed for the background model of \CUOREZ~\cite{Alduino:2016vtd} and became a standard for the background models of other \bb bolometric experiments~\cite{PhysRevLett.131.222501,PhysRevLett.123.262501,Augier2023BackgroundModelCUPID-Mo,PhysRevLett.131.162501}.}     
 Eventually, the activity of each background source is directly proportional to $\langle N_j \rangle$, that is the mode of the corresponding marginalized posterior distribution.
 
 \subsection{Diagonal-band method}
 \label{sec:M2}
 
 \begin{figure*}[tb]
    \centering
    \includegraphics[width=1.\columnwidth]{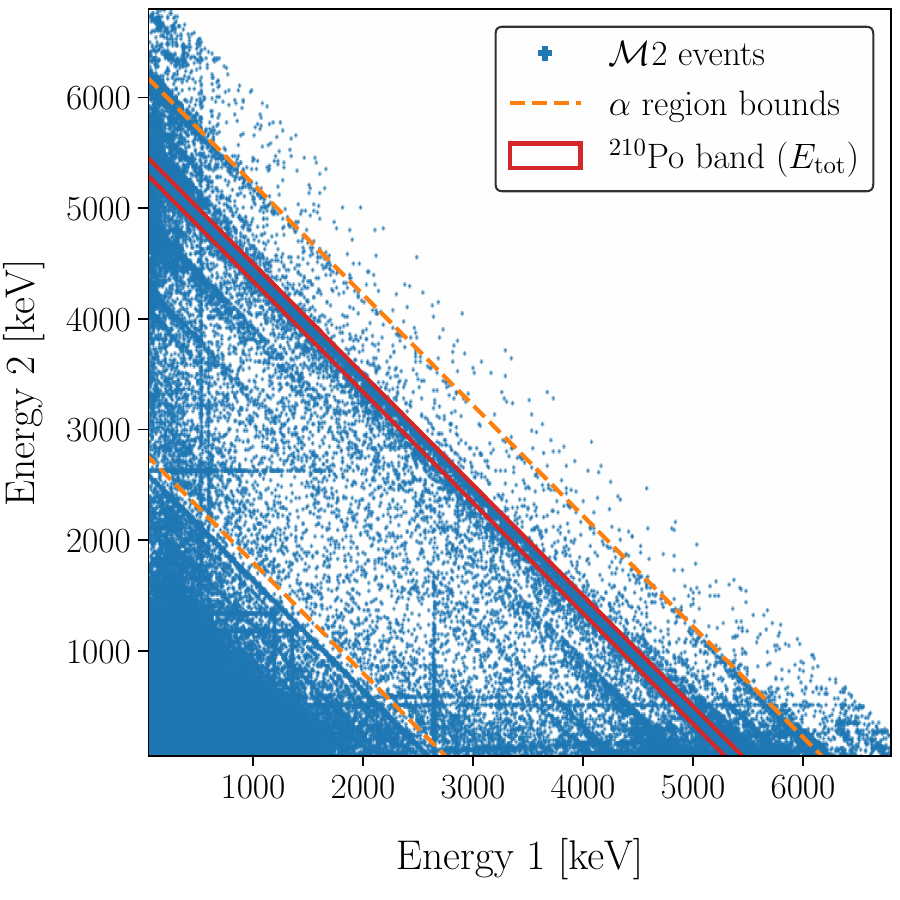}\includegraphics[width=1.\columnwidth]{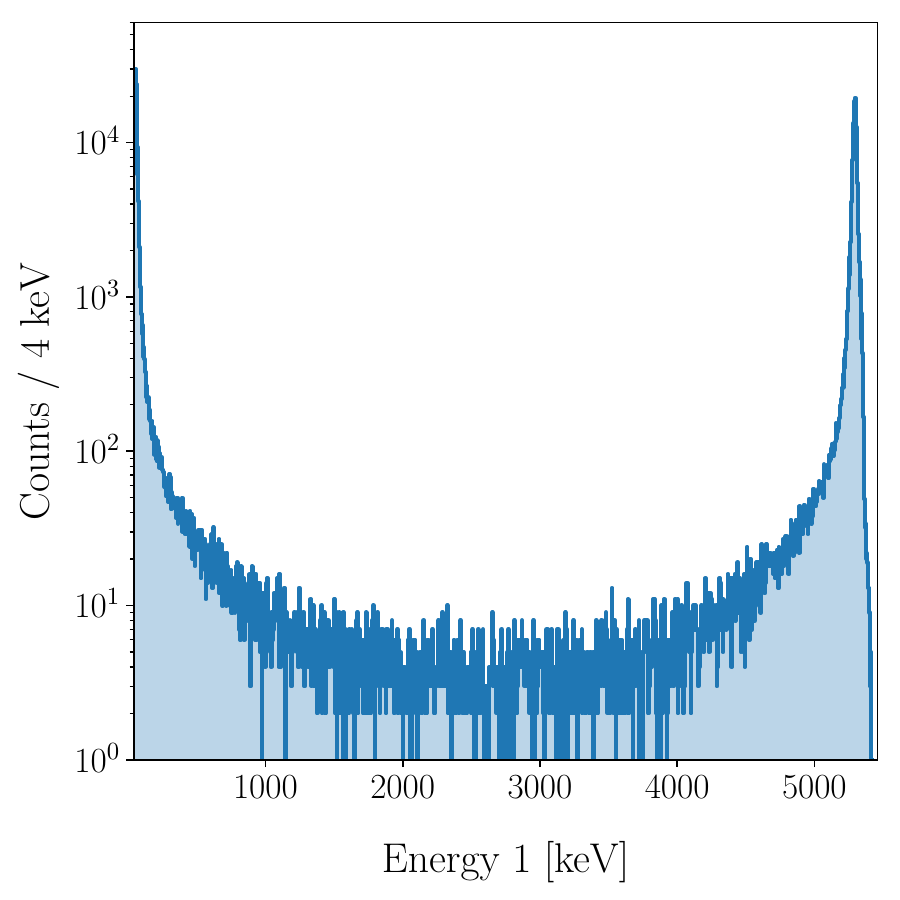}
    \caption{({\it Left}) Individual energy depositions in \MII data.
    The diagonal bands represent events with a constant total-energy shared between the two detectors involved.
    As an example, the region enclosed by the solid red line corresponds to total energies around the decay of \ce{^{210}Po} at $5.4$ MeV (summing to the \al energy and to the Q-value).
    The vertical and horizontal bands correspond to full-energy photo-peaks, where almost the totality of the \gm-ray energy is registered by one of the two detectors involved in the \MII event.
    ({\it Right}) Projection of the \ce{^{210}Po}-decays for \MII events (region inside the solid red line) onto the x-axis.  
    The peaks at $\sim 100$ keV $\sim 5300$ keV are due to the cases of \al particle and a recoiling nucleus detected in two different crystals; the region in-between is populated by events where the \al particle deposits a fraction of its energy in both crystals.
    }
    \label{fig:Scatter}
 \end{figure*}

 The fit of the \MII spectra allows us to exploit the collective information from the individual energy depositions of multi-crystals events to disentangle different background components.
 In our previous work~\cite{Alduino:2016vtd}, the data reconstruction was performed by simultaneously fitting the \MI energy spectrum together with \MII energy and \MII total-energy spectra, that are built by using the same data.
 In this analysis, we employ a novel {\it diagonal-band} method, in which we consider multiple uncorrelated \MII energy spectra, thus including precious physical information coming from the \MII total-energy spectrum while eliminating any redundancy.
 
 To describe this procedure, we initially focus on the \MII total-energy spectrum in the range $(2.7-6.8)$ MeV, which we refer to as the \al region, as most of the events are produced by \al decays. 
 This is shown in Fig.~\ref{fig:Scatter}, where we depict the \al region through the \MII energy-deposition scatter plot. The diagonal bands correspond to \MII events whose total-energy is about-constant, shared differently between the two involved crystals. 
 These bands are mainly caused by \al decays happening on the crystal surface, where one crystal detects the recoiling nucleus and possibly a fraction of the energy of the emitted \al particle, while the other crystal detects the remaining part of the \al-particle energy. 
 By analyzing the experimental \MII total-energy spectrum, we identify the peaks at the Q-values of the \al decays of \ce{^{232}Th} and \ce{^{238}U} chains (Table~\ref{tab:M2BandsList}). We then select the events laying inside the diagonal band associated to a certain total-energy peak and we project its content onto one of the two axes: the corresponding \MII energy spectra built from simulations are used for data reconstruction.
 As an example, the events due to the \al decay of \ce{^{210}Po} (red bordered region in Fig.~\ref{fig:Scatter}) have been selected by consider a total-energy between $5340$ and $5500$ keV. By repeating this method for a set of disjoint total-energy intervals, we generate a set of \MII spectra. The vast majority of the events within each \MII spectrum stems from the same contaminant in different locations. As a result, the degeneracy between background sources is mitigated and the correlations are reduced.

The selected intervals do not cover the whole \al region. Each selected diagonal band corresponds to a peak in the total-energy spectrum, while we exclude tails where the spectrum shape is not sufficiently well modeled in the simulation and regions in between peaks where no structure is present.  Ideally, the total energy of an \MII event corresponding to an \al decay should mainly fall within a narrow band around the transition Q-value, however we observe wide tails characterized by an increase in the count rate when approaching the Q-value. We explored two possible origins for these events:
\al particles or daughter nuclei scattering out of a crystal and ending up in a passive material; or \MII event not being due to an \al decay but rather some other multi-crystal event such as muon showers.  
However, the former has an extremely low probability of occurrence, while the latter requires a muon flux which is inconsistent with that observed in the experimental hall.
The energy ranges in question are predominantly populated by \al decays originating from the surface of detectors, which suggests that we are observing un-modeled surface effects. 
The fraction of events contained in these off-peak energy intervals is relatively small, $\sim 6\%$, therefore we exclude them from the reference background-model fit. 
To be conservative, we quote as a systematic uncertainty the difference between the fit with and without these regions included (Sec.~\ref{sec:systematics}).

Analogously to the \al case, we also split the region below $2.7$ MeV, which we refer to as the \gm region, into independent diagonal bands based on the peaks identified in the total-energy spectrum (Table~\ref{tab:M2BandsListGamma}). 
Since we can explain all the structures visible in the total-energy spectrum of the \gm region, here the bands cover the whole energy range.

 \subsection{Energy window and binning}
 \label{sec:EnergyWindowAndBinning}

 The fit window spans over the energy range $(0.2-6.8)$ MeV, apart from the \MII spectra of the \al region, for which we set the lower bound to $40$ keV in order to fully exploit the information on energy depositions of recoiling nuclei ($\sim 100$ keV).
 Widening the fit towards lower energies worsens the data-reconstruction quality because of possible missing or poorly-modeled background contributions and uncertainties on the detector response.
 At the same time, extending the upper bound to higher energies offers little benefit, as there are a few events and very few structures identifiable in the spectrum and our knowledge of energy-dependent quantities such as calibration and data-selection efficiencies becomes poor.
 
 We build the fit energy histograms using non-uniform energy bins.
 In the \gm region, around each identified line, we define single bins whose width is equivalent to $5$ times the energy resolution computed at the centroid of the peak. 
 We divide the regions that fall in between \gm-ray peaks into equally-sized bins, with a minimum size of $15$ keV for the \MI spectra and $40$ keV for the \MII spectra.
 This partition avoids systematic effects due to the energy calibration and peak-shape modeling.
 In the \al region, since we lack a satisfactory model of the off-peak tails, we manually select the bin edges to include all the counts from a specific spectral feature inside single wide bins. If required, we then merge adjacent bins to reach a minimum number of $50$ counts per bin. A summary of all the identified lines used to define the binning in both the \al and \gm regions is reported in App.~\ref{app:peaks}.
 
 \subsection{Prior selection}
 \label{sec:priors}
 
 As inputs to the Bayesian model, we make use of prior probability distributions which describe our existing knowledge of a specific contamination or, equivalently, fit parameter.
 When no \emph{a priori} information is applicable, we assume a uniform probability distribution ranging from zero to the maximum value that prevents the simulated component exceeding the data
 \footnote{In order to be more conservative, we compute maximum value of the prior by accounting for the statistical uncertainty of both data and \MC, therefore, fluctuating the data upwards by $2\sigma$ and the \MC downwards by $2\sigma$.}.
 Conversely, we make use of prior information the \CUOREZ background model~\cite{Alduino:2016vtd} or from independent sources, when available. 
 If a contamination has been measured, we assign a Gaussian probability with corresponding mean and standard deviation; in cases where there is only an upper limit, we take as prior an exponential distribution whose $90\%$ quantile matches the limit at $90\%$ C.\,L.\,. 

 For surface contamination of detector and near-detector elements, we observe contributions higher than what we would have inferred from \CUOREZ, mainly for the \ce{^{210}Pb} surface contamination of {\it\CuNOSV}. This is possibly due to recontamination of surfaces during the period when the towers were under storage. Therefore, in these cases we do not make use of priors based on external measurements.
 
 We set a prior for the simulation of cosmic muons as well. 
 Since the \MI and \MII energy spectra contain few muon-induced events, leaving the muon normalization unconstrained leads to a significant overestimation of the high-multiplicity spectra, which receive the main contribution from muons.
 Therefore, we first fit these high-multiplicity data and extract the prior of the corresponding normalization factor. The total muon flux we obtain is compatible with that measured by MACRO~\cite{Ambrosio:1995cx}.

 \subsection{Systematics}
 \label{sec:systematics}

 Inevitably, some of the assumptions made constructing the background model are a potential source of systematic uncertainty.
 Examples include our parametrization of the detector response to \al particles with a \QF, the assumption that the background contributions are uniformly distributed in each volume, and the assumption that activities are constant over time.

 We identify different classes of systematic uncertainties which share a common cause in the fit specifications and we estimate their impact on the background model results by repeating the fit while varying these specifications.
 We then compare the result with the default fit and use the difference we observe to quantify the associated systematic uncertainty. 
 
 In particular, we take into account the following effects:
 \begin{description}
  \item[{\it Binning}] we consider different (constant) bin-widths for the \gm region and we perform a separate fit where we include the off-peak intervals in the \MII spectra associated with the \al region;
  \item[{\it Energy threshold}] we move the low-energy cut off of the fit in the range ($150-250$) keV; 
  \item[{\it Geometry}] we individually fit each floor and each tower of \CUORE;
  \item[{\it Dataset}] we individually fit each of the $15$ datasets used in this analysis;
  \item[{\it \ce{^{90}Sr}}] we repeat the fit adding a contamination of \ce{^{90}Sr} in the \ce{TeO_{2}} (see App.~\ref{app:TeO2OtherContaminants}).
 \end{description}

We note that for the {\it Binning} and {\it Energy threshold} classes, exploring the systematic variation of the results requires fitting partially or completely overlapping set of data. Therefore, when computing the corresponding uncertainty, we conservatively pick the largest deviation from the default fit.
For the {\it Geometry} and {\it Dataset} classes, where the fit is repeated on independent subdivisions of the data, we quantify the associated systematic as the average distance of the results from the default fit, weighted for the inverse-square of the posterior width.
In both cases, if the reference fit converges to a value different from zero, we then subtract in quadrature the average statistical uncertainty of the single fits; conversely, if the reference fit is compatible with zero, we quote a 90\% credibility interval (C.I.) limit.

 \section{Results}
 
 \begin{figure*}[t]
  \centering
  \includegraphics[width=.95\textwidth]{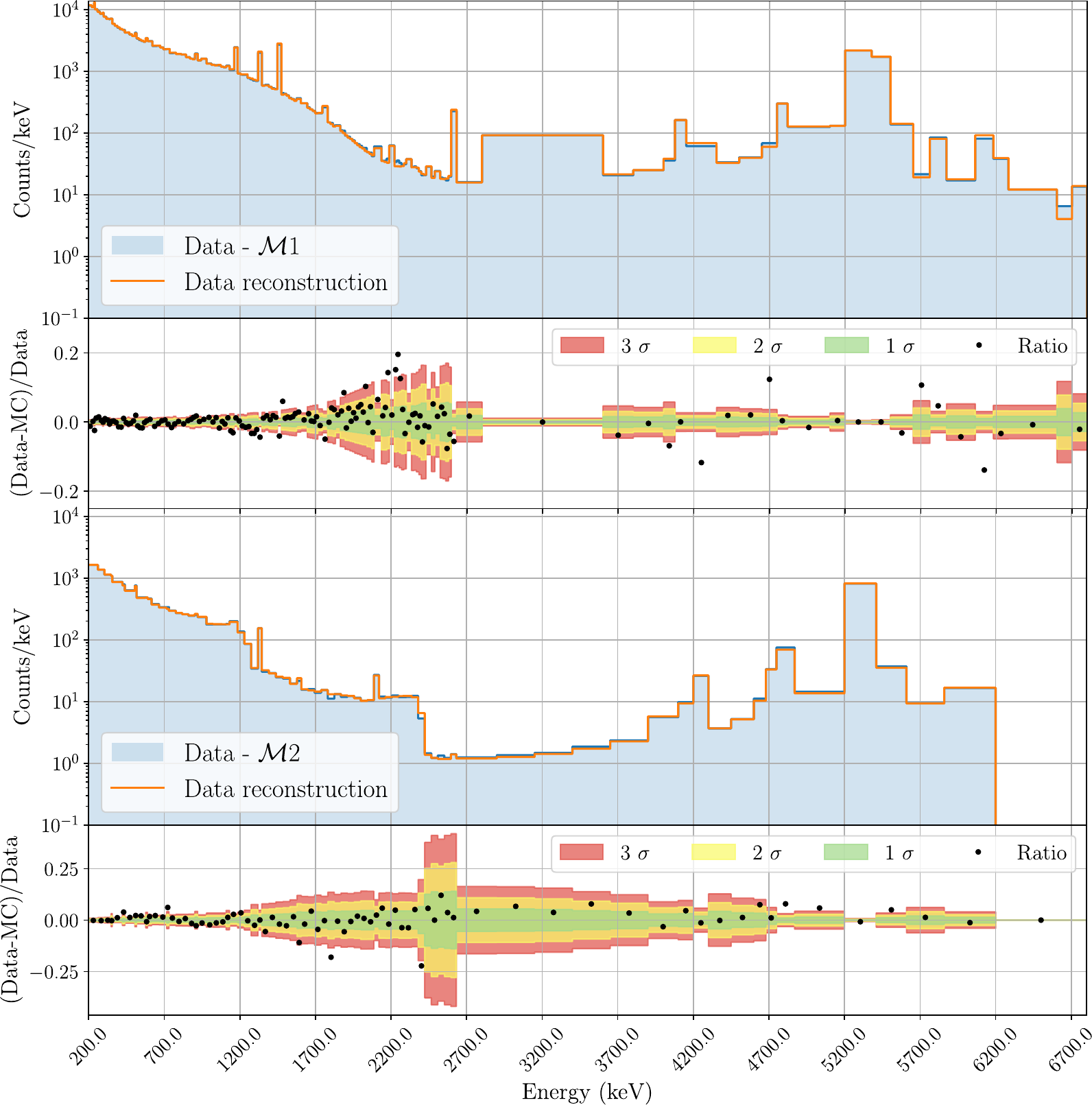}
  \caption{\emph{(Top panels)} Reconstruction of the \MI events and residuals.
   \emph{(Bottom panels)} Cumulative spectrum of the individual \MII events and residuals. 
   The residuals are quoted as fractions of data, with the statistical uncertainties represented as $1\sigma, 2\sigma$ and $3\sigma$ standard-deviation bands.
   The \MII spectrum ends at $6200$ keV because this is the maximum energy for the events in the highest total-energy band taken into account.}
  \label{fig:M1M2Reconstructions}
 \end{figure*}
 
\begin{figure*}[t]
  \centering
  \includegraphics[width=1.\textwidth]{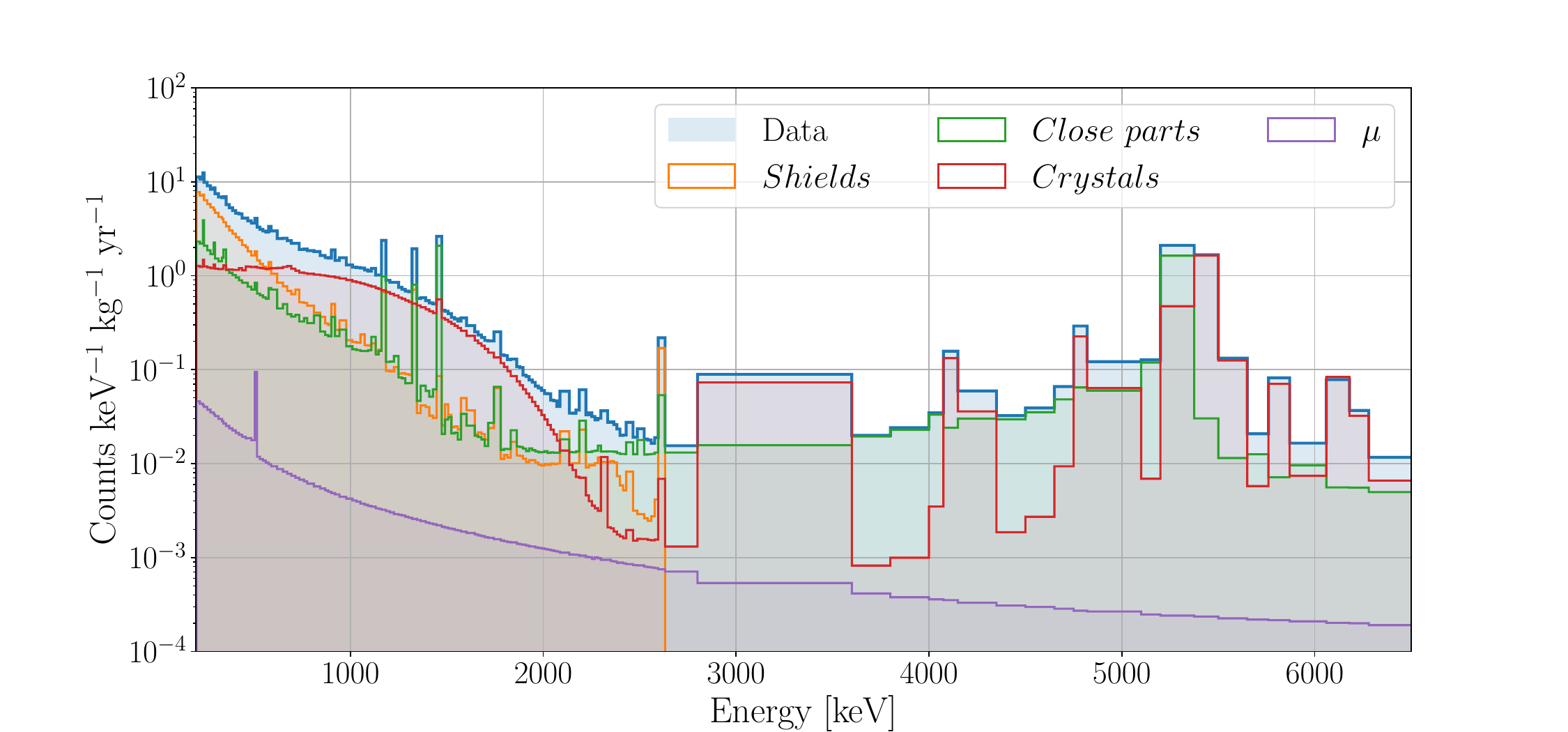}
  \caption{Decomposition of the CUORE data by source location. The {\it{Shields}} group includes the volumes {\it\ModPbOut}, {\it\CuOFEOut}, {\it\ModPbIn}, {\it\RomanPb} and {\it\CuOFEIn}.}
  \label{fig:CUOREBackgroundDecomposition}
 \end{figure*} 
 
 In Fig.~\ref{fig:M1M2Reconstructions}, we show the reconstruction of the \CUORE data with the background model.
 The top plot contains the fit of \MI events, while the bottom one depicts the projections of the \MII energy histograms used for the fit onto a single spectrum.
 In general, we observe a good agreement between the fit and the observed data and we find only a few bins in the \MI reconstruction which show significant residuals. These are mainly in the \al region, where our understanding of the detector response is incomplete and we cannot describe the peak shape, and around $2.2$ MeV, where we observe an excess in the counts.

 By grouping the different contributions according to the volumes used to simulate the \CUORE geometry, we obtain the spectrum decomposition of Fig.~\ref{fig:CUOREBackgroundDecomposition}.
 In the low-energy region until $400$ keV, the leading contribution to the background is represented by Bremsstrahlung photons following the \bt decay of \ce{^{210}Bi}, mainly from the {\it\ModPbIn}, {\it\HEX} and {\it\MX} volumes.
 Between $500$ keV and $2$ MeV, the spectrum is dominated by the \bbvv of \ce{^{130}Te}.
 The remaining part of the \gm region mostly sees a continuum from the very-shallow contamination of \ce{^{232}Th}, \ce{^{238}U} and \ce{^{210}Pb} mainly from {\it\CuNOSV}.
 The \al region contains the contribution of multiple \al emitters from the \ce{U} and \ce{Th} decay chains due to contaminants shared between {\it\xtals}, mostly producing full-energy peaks, and {\it\CuNOSV}, giving a continuum due to degraded \al-events from copper.
 This results in a flat spectrum until $4$ MeV, below which the line of \ce{^{190}Pt} from the bulk contamination of the crystals is the only clear signature present. Above we see multiple peaks, the most prominent coming from the decay of \ce{^{210}Po}.

 The complete list of reconstructed activities is reported in Tables~\ref{tab:Results_Bulk_contamination} and \ref{tab:Results_Surface_contamination} for the bulk and surface contributions, respectively.
 Each activity is provided either as a value, with associated statistical uncertainty, or as a limit. In the former case, we extract the mode of the posterior probability distribution and take the narrowest $68\%$-interval containing the mode to estimate the uncertainty; in the latter case, we take the $90\%$ quantile of the posterior.
 We compute systematic uncertainty ranges for each class according to the procedure described in the previous section and the systematic uncertainty reported is the narrowest interval that contains these bands.
 
We tested the stability of the results by repeating the analysis without informative priors, except for the muon normalization. We observe no noteworthy differences with respect to the reference fit. When considering both statistical and systematic uncertainties, all the activities coming from the reference fit are compatible with the considered prior knowledge. The detailed model coupled with the high collected statistics allows \CUORE to measure activities down to $\sim 10$ nBq kg$^{-1}$ for bulk and $\sim 0.1$ nBq cm$^{-2}$ for surface contamination, reaching the highest sensitivity for very-close sources. 

In general, statistical uncertainties are low, on the order of 15\% or less. 
We observe correlations mainly among \emph{\CuNOSV}, \emph{\CuOFEIn} and \emph{\RomanPb} for \ce{^{60}Co}, \ce{^{40}K} and the natural decay chains. This is due to the very degenerate contribution of these sources and limited prior knowledge.

 Systematic uncertainties caused by varying the \emph{binning} or \emph{energy threshold} are subdominant. On the other hand, the biggest variations are observed in the \emph{geometry} and \emph{dataset} classes. 
 In the \emph{geometry} class, some variations in fit results were expected while others were not. For example, \ce{^{190}Pt} is accidentally included during the crystal growth in the form of microscopic clusters and this naturally leads to sparse and unpredictable contamination in the crystals. On the other hand, we observe that the activities of the most superficial \ce{^{210}Pb} in both \emph{\CuNOSV} and \emph{\xtals} show a very scattered distribution when fitting single floors and towers. This is likely due to different exposure to \ce{^{222}Rn} during the transportation and storage phases before the experiment.
 Another source of geometric variation shows up as monotonic trends in the activity of \ce{^{210}Pb} in \ModPbIn, \HEX and \MX with respect to top-bottom slices of the detector. 
 This can be connected with the asymmetrical arrangement of the cryostat elements, concentrated in the upper part (\TSP and \ModPbIn) and around the structure (mainly \MX and \HEX) which is correctly rendered in the simulation. This points to possible dis-uniformities in the activity of the innermost shields.

\begin{figure}[t]
  \centering
  \includegraphics[width=1.\columnwidth]{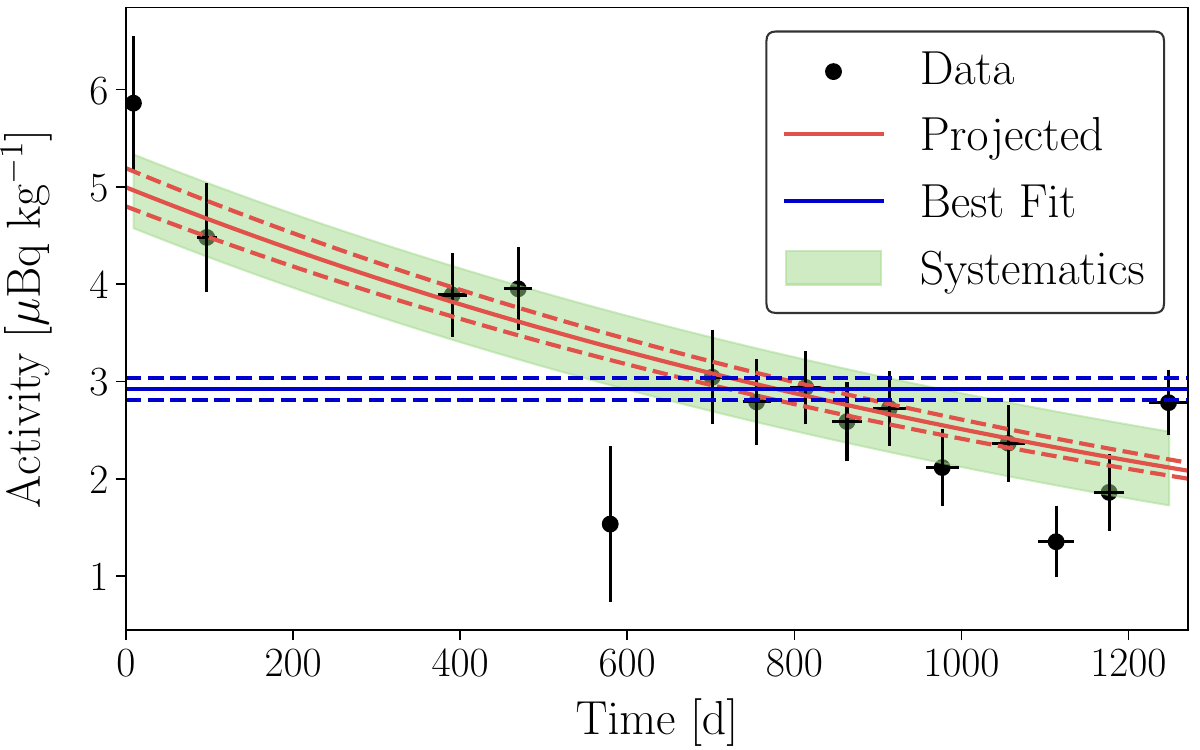}
  \caption{Activity of \ce{^{125}Sb} in the {\it \xtals} volume extracted from the individual-dataset fits (black dots).
  The blue lines (solid and dashed) depict the mode and the statistical uncertainty of the activity from the best fit, where all the dataset are grouped together. The red line is the projected time-dependent activity, computed by assuming the best fit to be the integral average along time of a decaying activity with the tabulated half-life of 2.76 yr.
  The green band shows the systematic uncertainty coming from the comparison between each individual-dataset fit and the time-dependent activity obtained by the reference fit.}
  \label{fig:54MnVsDS}
 \end{figure}

 Turning to the \emph{dataset} systematic class, we observe relatively small time-related variations in some activities of isotopes with very long half-life or in equilibrium with their progenitors. However, expected dataset-dependent effects, such as channel status and analysis efficiencies, are accounted for during the \ares processing. We therefore fold these into the associated uncertainty as described above. 
 Conversely, there are cases where a time-dependent activity is expected, as for the cosmogenic activation isotopes \ce{^{125}Sb}, \ce{^{54}Mn}, and \ce{^{60}Co}, which have half-lives comparable to the data-collection time of \CUORE.
 For such cases, we compute the systematic uncertainty differently, by utilizing the distance of the single fits with respect to the decay trend computed from the reference fit value, as exemplified in Fig.~\ref{fig:54MnVsDS}. The starting specific activity for the decay trend is determined from reference-fit specific activity, as it represents the time-integrated counts per dataset over the total exposure of the datasets while the specific decay half-life is fixed to tabulated values.
 This type of analysis is important to verify {\it a-posteriori} if the storage time of the components was suitable and to study the continuous material activation underground which gives a time-constant contribution. 
 In addition to the cosmogenic isotopes mentioned above, \ce{^{210}Pb} in both \textit{\xtals} and \textit{\CuNOSV} shows a clear decay in time that matches the expected half-life of \ce{^{210}Po}, that is 138.4 days. The component not in equilibrium with \ce{^{210}Pb} completely decayed within the first two datasets.
 
\begin{figure}[t]
  \centering
  \includegraphics[width=1.\columnwidth]{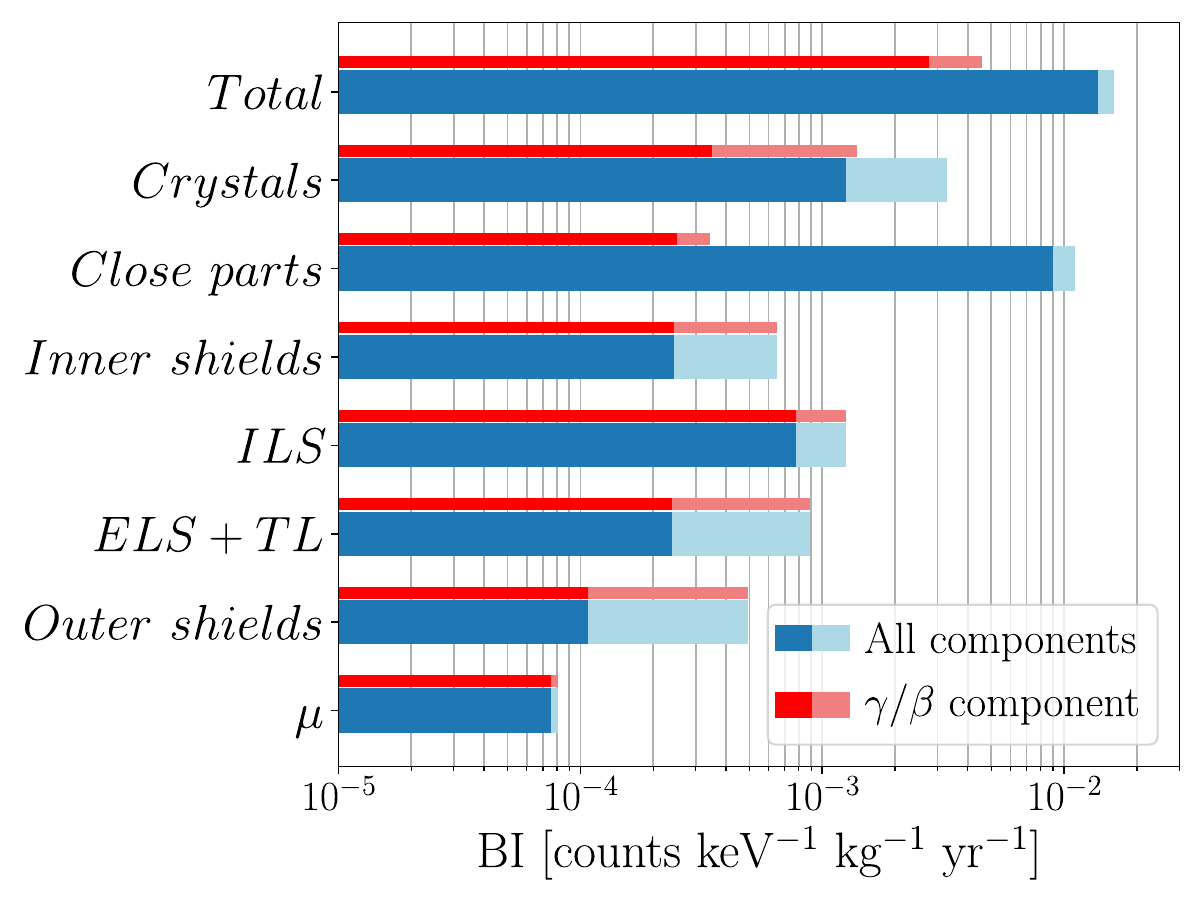}
  \caption{BI components obtained from the full set of fits used to extract the systematic uncertainty. The light (dark) blue band represent the 16\% (84\%) quantile of the BI distribution obtained. The red bands indicate the corresponding fraction due to \bt and \gm events.}
\label{fig:CUOREBackgroundsROI}
\end{figure}

 Another interesting result from the background model is the set of components contributing to background in the region of interest for the \bb search near $Q_{\beta\beta}=
 2526.97(23)$ keV~\cite{RAHAMAN2011412}. 
 It has to be noted that the cuts adopted for the \CUORE background model differ from those of the \bb studies. In this work, we set a less stringent pulse-shape cut and a different coincidence window. These data-selections have been specifically optimized for this analysis, which covers the whole energy range of the detector and it is not limited to a narrow Region Of Interest (ROI) around the $Q_{\beta\beta}$.
 In order to extract the ROI background index (BI), 
 we apply to the MC simulations of the background sources the same pulse-ts and the same coincidence window used for the \bb analysis.
 \newline
 The result is shown in Fig.~\ref{fig:CUOREBackgroundsROI} and has been derived by evaluating the integral of all the background components in the ROI, repeating this procedure for all the fits utilized for the systematic studies. The light blue and blue bars refer to the 16\% and 84\% quantiles of the resulting distributions, respectively. Their red counterparts quote the background induced by only considering \gm and \bt particles.

 The total BI$=1.48_{-0.10}^{+0.12}\times 10^{-2}$ {\mbox counts keV$^{-1}$ kg$^{-1}$ yr$^{-1}$}, where the value and the uncertainty come from the average BI and the [0.16\%, 0.84\%] quantiles of the full set of fits used to study the systematic uncertainties, respectively. 

 In particular, we find that approximately 75\% of the $BI$ is attributable to \al events, mainly coming from the {\it \CuNOSV}, where we define an \al event to be one where at least 90\% of the event energy came from an \al decay. This is consistent with the BI obtained with the \bb fit applied on the same set of data~\cite{CUORE:2021mvw}, that is $1.49\pm 0.04\times 10^{-2}$ {\mbox counts keV$^{-1}$ kg$^{-1}$ yr$^{-1}$}. 
 Upon comparing these results with the CUORE background projections~\cite{Alduino:2017qet} we observe no noteworthy difference except for {\it \CuNOSV}, which shows a $\sim 15\%$ higher contribution to the BI because of the aforementioned possible recontamination. While the bulk contamination is in line with the projections, surface contamination is higher than expected. Indeed, bulk activities can be easily measured with high precision; however, the reliability of surface contamination estimates is hindered by challenging measurements and potential risk of recontamination during the storage and commissioning phases.
 
  \begin{center}
  \begin{table*}[t]
   \caption{Activities of the bulk contamination and muon flux.
    For each volume, the individual contributions are listed; where a single nuclide is reported we refer to its full decay chain.
    When present, we include our prior knowledge from either \CUOREZ~\cite{Alduino:2016vtd}, from radio-assay measurements (HPGe or Neutron-Activation Analysis).
    The posterior modes, i.\,e.\ the fit results, are quoted together with their statistical uncertainty and the largest associated systematic error, which is implicitly expressed with the same order of magnitude of the reference value. When the mode is compatible with zero, we quote the 90\% C.I. as a limit on the activity.
    The activity of \ce{^{235}U} is fixed with respect to that of \ce{^{238}U} by their natural activity ratio.
    For the specific activity of \ce{^{ 40} K} in tower 12 see App.~\ref{app:K_t12}.}
   \bigskip
   \begin{tabular}{l r r r r c}
    Volume  &\hphantom{cccc}Contaminant  &\multicolumn{2}{r}{\hspace{60pt}Prior $[\text{Bq kg}^{-1}]$}   &\hspace{15pt}Mode/Limit $[\text{Bq kg}^{-1}]$ &\hspace{10pt}Systematic \\[2pt]
    \hline  \\[-10pt] {\it \xtals} \\ \cline{1-1} \\[-15pt]
    &	\ce{^{130}Te} \bbvv 	&		                    &	 	    &	$(3.03 \pm 0.01) \times 10^{-5}$ 	&	$_{-0.17}^{+0.11}$  \\
    &	\ce{^{232}Th } 	&	$<1.2\times 10^{-7}$	&	\CUOREZ	&	$(2.75 \pm 0.05) \times 10^{-7}$ 	&	$_{-1.47}^{+0.85}$  \\
    &	\ce{^{228}Ra \to ^{208}Pb} 	&	$<7.5\times 10^{-8}$	&	\CUOREZ	&	$(1.19 \pm 0.04) \times 10^{-7}$ 	&	$_{-1.16}^{+0.2}$  \\
    &	\ce{^{238}U \to ^{230}Th}	&$<3.6\times 10^{-8}$	&	\CUOREZ	&	$<6.36 \times 10^{-10}$ 	&	  \\
    &	\ce{^{230}Th } 	&	$(2.8  \pm 0.3)  \times 10^{-7}$ 	&	\CUOREZ	&	$(3.85 \pm 0.06) \times 10^{-7}$ 	&	$_{-1.3}^{+0.26}$  \\
    &	\ce{^{226}Ra \to ^{210}Pb} 	&	$<2.2\times 10^{-8}$	&	\CUOREZ	&	$<4.63 \times 10^{-10}$ 	&	  \\
    &	\ce{^{210}Pb} 	&	$(1.37 \pm 0.83) \times 10^{-6}$	&	\CUOREZ	&	$(1.55 \pm 0.02) \times 10^{-6}$ 	&	$_{-1.48}^{+0.44}$  \\
    &	\ce{^{235}U \to ^{231}Pa} 	&		                    &	 	    &	$<2.92 \times 10^{-11}$ 	&	  \\
    &	\ce{^{231}Pa \to ^{207}Pb} 	&		                    &	 	    &	$<9.06 \times 10^{-10}$ 	&	  \\
    &	\ce{^{190}Pt} 	&	\hphantom{cc}$(1.95 \pm 0.05) \times 10^{-6}$	&	\hphantom{c}\CUOREZ	&	$(1.93 \pm 0.01) \times 10^{-6}$ 	&	$_{-0.3}^{+0.29}$  \\
    &	\ce{^{147}Sm} 	&		                    &	 	    &	$(1.09 \pm 0.12) \times 10^{-8}$ 	&	$_{-0.58}^{+0.67}$  \\
    &	\ce{^{125}Sb} 	&		                    &	 	    &	$(2.93 \pm 0.11) \times 10^{-6}$ 	&	$_{-1.44}^{+2.42}$  \\
    &	\ce{^{110m}Ag} 	&		                    &	 	    &	$(9.06 \pm 2.44) \times 10^{-8}$ 	&	$_{-2.45}^{+62.58}$  \\
    &	\ce{^{108m}Ag} 	&		                    &	 	    &	$(6.02 \pm 1.08) \times 10^{-8}$ 	&	$_{-2.66}^{+2.61}$  \\
    &	\ce{^{60}Co} 	&	$(3.0  \pm 1.4)  \times 10^{-7}$	&	\CUOREZ	&	$(1.86 \pm 1.22) \times 10^{-8}$ 	&	$_{ }^{+4.21}$  \\
    &	\ce{^{40}K} (no Tower 12) 	&	$< 8.2\times 10^{-6}$	&	\CUOREZ	&	$(4.30 \pm 0.12) \times 10^{-6}$ 	&	$_{-1.11}^{+2.62}$  \\
    &	\ce{^{40}K} (Tower 12) 	&     &	&	$(2.45 \pm 0.68) \times 10^{-5}$ 	&	$_{-0.63}^{+1.49}$  \\[-3pt]
    {\it \CuNOSV} \\ \cline{1-1} \\[-15pt]
    &	\ce{^{232}Th} 	&	$<2.1\times 10^{-6}$	&	\CUOREZ	&	$<3.88 \times 10^{-7}$ 	&	  \\
    &	\ce{^{238}U} 	&	$<2.2\times 10^{-5}$	&	\CUOREZ	&	$<4.73 \times 10^{-7}$ 	&	  \\
    &	\ce{^{235}U} 	&	&	&	$<2.17 \times 10^{-8}$ 	&	  \\
    &	\ce{^{137}Cs} 	&	$<2.2\times 10^{-5}$	&	HPGe	&	$(1.25 \pm 0.24) \times 10^{-6}$ 	&	$_{-0.71}^{ }$  \\
    &	\ce{^{60}Co} 	&	$<2.5\times 10^{-5}$	&	HPGe	&	$(2.04 \pm 0.03) \times 10^{-5}$ 	&	$_{-0.39}^{+0.32}$  \\
    &	\ce{^{54}Mn} 	&	$<3.1\times 10^{-5}$ 	&	HPGe	&	$(2.29 \pm 0.33) \times 10^{-6}$ 	&	$_{-1.93}^{+2.63}$  \\
    &	\ce{^{40}K} 	&		                    &		    &	$(4.42 \pm 0.06) \times 10^{-4}$ 	&	$_{-1.06}^{ }$  \\
    {\it \CuOFEIn} \\ \cline{1-1} \\[-15pt]
    &	\ce{^{232}Th} 	&	$<6.4\times 10^{-5}$	&	HPGe	&	$(4.10 \pm 0.39) \times 10^{-5}$ 	&	$_{-2.54}^{+1.92}$  \\
    &	\ce{^{238}U} 	&	$<5.4\times 10^{-5}$	&	HPGe	&	$(7.71 \pm 5.03) \times 10^{-6}$ 	&	$_{ }^{+16.51}$  \\
    &	\ce{^{137}Cs} 	&		                    &	 	    &	$<1.92 \times 10^{-6}$ 	&	  \\
    &	\ce{^{60}Co} 	&	$<2.4\times 10^{-5}$	&	HPGe	&	$(1.46 \pm 0.19) \times 10^{-5}$ 	&	$_{-1.44}^{+4.89}$  \\
    &	\ce{^{54}Mn} 	&		                    &	 	    &	$<3.71 \times 10^{-6}$ 	&	  \\
    &	\ce{^{40}K} 	&	$<6.7\times 10^{-4}$	&	HPGe	&	$<3.48 \times 10^{-5}$ 	&	  \\
    {\it \CuOFEOut}\hphantom{cc} \\ \cline{1-1} \\[-15pt]
    &	\ce{^{232}Th} 	&		                    &	 	    &	$<2.45 \times 10^{-5}$ 	&	  \\
    &	\ce{^{238}U} 	&		                    &	 	    &	$<4.02 \times 10^{-5}$ 	&	  \\
    &	\ce{^{137}Cs} 	&		                    &	 	    &	$<7.33 \times 10^{-4}$ 	&	  \\
    &	\ce{^{60}Co} 	&		                    &	 	    &	$(1.45 \pm 0.04) \times 10^{-3}$ 	&	$_{-0.87}^{+0.29}$  \\
    &	\ce{^{54}Mn} 	&		                    &	 	    &	$<2.14 \times 10^{-4}$ 	&	  \\
    &	\ce{^{40}K} 	&		                    &	 	    &	$<8.61 \times 10^{-4}$ 	&	  \\
    {\it \RomanPb} \\ \cline{1-1} \\[-15pt]
    &	\ce{^{232}Th} 	&	$ (3.9 \pm 2.2)  \times 10^{-5}$	&	\CUOREZ	&	$(1.70 \pm 0.22) \times 10^{-5}$ 	&	$_{-0.8}^{+0.62}$  \\
    &	\ce{^{238}U} 	&	$ (2.7 \pm 1.0)  \times 10^{-5}$	&	\CUOREZ	&	$<1.61 \times 10^{-6}$ 	&	 $<11.44$ \\
    &	\ce{^{108m}Ag} 	&		                    &	 	    &	$(7.99 \pm 0.78) \times 10^{-6}$ 	&	$_{-3.72}^{+2.62}$  \\
    &	\ce{^{40}K} 	&	&	&	$<3.87 \times 10^{-5}$ 	&	 $<18.58$ \\
    {\it \ModPbIn} \\ \cline{1-1} \\[-15pt]
 	&	\ce{^{232}Th} 	&	&  	&	$(3.06 \pm 1.47) \times 10^{-4}$ 	&	$_{-2.74}^{+22.95}$  \\
 	&	\ce{^{238}U} 	&	$< 1.1           \times 10^{-3}$	&	HPGe	&	$(3.45 \pm 0.36) \times 10^{-3}$ 	&	$_{-3.44}^{ }$  \\ 
 	&	\ce{^{210}Bi \to ^{206}Pb} 	&		&	 	&	$(1.61 \pm 0.02) \times 10^{+2}$ 	&	$_{-0.41}^{+0.51}$  \\
    &   \ce{^{40}K} &$< 7.6\times 10^{-3}$	&	HPGe   &$(3.74 \pm 2.64) \times 10^{-3}$ &$_{-3.01}^{+7.49}$  \\[-3pt]
  {\it \ModPbOut} \\ \cline{1-1} \\[-15pt]
    &	\ce{^{210}Bi} 	&              &	 	    &	$(3.31 \pm 0.14) \times 10^{+2}$ 	&	$_{-1.86}^{+1.35}$  \\
    &	\ce{^{207}Bi} 	&              &	 	    &	$(2.29 \pm 0.20) \times 10^{-3}$ 	&	$_{-1.47}^{+1.21}$  \\
       \end{tabular}
   \label{tab:Results_Bulk_contamination}
  \end{table*}
 \end{center}

  \begin{center}
  \begin{table*}[t]
   \caption{Activities of the surface contamination.
    For each volume, the individual contributions at different depths are listed; where a single nuclide is reported, we refer to its full decay chain.
    The posteriors modes, i.\,e.\ the fit results, are quoted together with their statistical uncertainty and the largest associated systematic error, which is implicitly expressed with the same order of magnitude of the reference value. When the mode is compatible with zero, we quote the 90\% C.I. as a limit on the activity.
    The activity of \ce{^{235}U} is fixed with respect to that of \ce{^{238}U} by their natural activity ratio.}
   \bigskip
   \begin{tabular}{l r rl r c}
     Volume          &\hspace{15pt}Contaminant   &\multicolumn{2}{r}{\hspace{10pt}Depth $[\mu\text{m}]$}   &\hspace{15pt}Mode/Limit $[\text{Bq cm}^{-2}]$ &\hspace{10pt}Systematic\\[2pt]
    \hline  \\[-10pt] {\it \xtals} \\ \cline{1-1} \\[-15pt]
 	&	\ce{^{210}Pb} 	&	$0$	&   \!\!$.001$	&	$(7.32 \pm 0.02) \times 10^{-8}$ 	&	$_{-3.23}^{+4.98}$  \\
 	&	\ce{^{232}Th} 	&	$0$	&   \!\!$.01$	&	$(3.10 \pm 0.14) \times 10^{-10}$ 	&	$_{-2.98}^{+0.2}$  \\
    &	\ce{^{228}Ra \to ^{208}Pb} 	&	$0$	&   \!\!$.01$	&	$(1.10 \pm 0.03) \times 10^{-9}$ 	&	$_{-0.19}^{+0.69}$  \\
 	&	\ce{^{238}U \to ^{230}Th} 	&	$0$	&   \!\!$.01$	&	$(3.79 \pm 0.06) \times 10^{-7}$ 	&	$_{-2.16}^{}$  \\    
    &	\ce{^{230}Th} 	&	$0$	&   \!\!$.01$	&	$(8.22 \pm 0.32) \times 10^{-10}$ 	&	$_{-0.4}^{+13.51}$  \\
    &	\ce{^{226}Ra \to ^{210}Pb} 	&	$0$	&   \!\!$.01$	&	$(2.56 \pm 0.04) \times 10^{-9}$ 	&	$_{-1.12}^{+1.52}$  \\
 	&	\ce{^{235}U \to ^{231}Pa} 	&	$0$	&   \!\!$.01$	&	$(1.74 \pm 0.03) \times 10^{-8}$ 	&	$_{-0.99}^{}$  \\    
 	&	\ce{^{231}Pa \to ^{207}Pb} 	&	$0$	&   \!\!$.01$	&	$(1.05 \pm 0.34) \times 10^{-10}$ 	&	$_{-0.66}^{+1.07}$  \\
    &	\ce{^{232}Th} 	&	$0$	&   \!\!$.1$	&	$(3.21 \pm 1.52) \times 10^{-11}$ 	&	$_{}^{+3.21}$  \\
 	&	\ce{^{228}Ra \to ^{208}Pb} 	&	$0$	&   \!\!$.1$	&	$(5.34 \pm 0.34) \times 10^{-10}$ 	&	$_{-5.27}^{}$  \\
 	&	\ce{^{238}U \to ^{230}Th} 	&	$0$	&   \!\!$.1$	&	$(1.83 \pm 0.53) \times 10^{-8}$ 	&	$_{-1.67}^{+7.3}$  \\
 	&	\ce{^{230}Th} 	&	$0$	&   \!\!$.1$	&	$(8.64 \pm 2.56) \times 10^{-11}$ 	&	$_{-3.98}^{+7.75}$  \\
 	&	\ce{^{226}Ra \to ^{210}Pb} 	&	$0$	&   \!\!$.1$	&	$(9.10 \pm 0.40) \times 10^{-10}$ 	&	$_{-8.71}^{+1.31}$  \\
    &	\ce{^{210}Pb} 	&	$0$	&   \!\!$.1$	&	$(1.31 \pm 0.01) \times 10^{-8}$ 	&	$_{-0.17}^{+0.29}$  \\
 	&	\ce{^{235}U \to ^{231}Pa} 	&	$0$	&   \!\!$.1$	&	$(8.42 \pm 2.39) \times 10^{-10}$ 	&	$_{-7.69}^{+33.59}$  \\
 	&	\ce{^{231}Pa \to ^{207}Pb} 	&	$0$	&   \!\!$.1$	&	$<6.06 \times 10^{-11}$ 	&	  \\
   	&	\ce{^{232}Th} 	&	$1$	&   &	$(7.77 \pm 1.74) \times 10^{-11}$ 	&	$_{-3.81}^{}$  \\
    &	\ce{^{228}Ra \to ^{208}Pb} 	&	$1$	&   &	$(1.86 \pm 0.19) \times 10^{-10}$ 	&	$_{-1.06}^{+10.17}$  \\
 	&	\ce{^{238}U \to ^{230}Th} 	&	$1$	&   &	$(5.68 \pm 0.28) \times 10^{-8}$ 	&	$_{-2.22}^{+1.01}$  \\
 	&	\ce{^{230}Th} 	&	$1$	&   &	$(9.32 \pm 1.84) \times 10^{-11}$ 	&	$_{-5.25}^{+18.73}$  \\
 	&	\ce{^{226}Ra \to ^{210}Pb} 	&	$1$	&   &	$(3.08 \pm 0.15) \times 10^{-10}$ 	&	$_{-2.58}^{+1.41}$  \\
 	&	\ce{^{210}Pb} 	&	$1$	&   &	$(5.15 \pm 0.10) \times 10^{-9}$ 	&	$_{-0.94}^{+0.7}$  \\
 	&	\ce{^{235}U \to ^{231}Pa} 	&	$1$	&   &	$(2.61 \pm 0.13) \times 10^{-9}$ 	&	$_{-1.02}^{+0.46}$  \\
 	&	\ce{^{231}Pa \to ^{207}Pb} 	&	$1$	&   &	$<2.23 \times 10^{-11}$ 	&	  \\
    &	\ce{^{232}Th} 	&	$10$	&   &	$(1.18 \pm 0.28) \times 10^{-10}$ 	&	$_{}^{+7.12}$  \\
 	&	\ce{^{228}Ra \to ^{208}Pb} 	&	$10$	&   &	$(3.29 \pm 1.27) \times 10^{-11}$ 	&	$_{}^{+61.54}$  \\
 	&	\ce{^{238}U \to ^{230}Th} 	&	$10$	&   &	$<3.97 \times 10^{-9}$ 	&	  \\
 	&	\ce{^{230}Th} 	&	$10$	&   &	$(2.17 \pm 0.25) \times 10^{-10}$ 	&	$_{-0.78}^{+5.95}$  \\
 	&	\ce{^{226}Ra \to ^{210}Pb} 	&	$10$	&   &	$(1.82 \pm 0.86) \times 10^{-11}$ 	&	$_{-1.46}^{+10.24}$  \\
    &	\ce{^{210}Pb} 	&	$10$	&   &	$(2.23 \pm 0.09) \times 10^{-9}$ 	&	$_{-2.18}^{+2.48}$  \\
 	&	\ce{^{235}U \to ^{231}Pa} 	&	$10$	&   &	$<1.83 \times 10^{-10}$ 	&	  \\
 	&	\ce{^{231}Pa \to ^{207}Pb} 	&	$10$	&   &	$<1.37 \times 10^{-11}$ 	&	  \\

    {\it \CuNOSV}\hphantom{cc} \\ \cline{1-1} \\[-15pt]
 	&	\ce{^{232}Th} 	&	$0$	&   \!\!$.01$	&	$(1.35 \pm 0.06) \times 10^{-9}$ 	&	$_{-0.51}^{+0.51}$  \\
 	&	\ce{^{238}U} 	&	$0$	&   \!\!$.01$	&	$(1.24 \pm 0.07) \times 10^{-9}$ 	&	$_{-0.68}^{+0.44}$  \\
 	&	\ce{^{210}Pb} 	&	$0$	&   \!\!$.01$	&	$(3.40 \pm 0.02) \times 10^{-7}$ 	&	$_{-0.96}^{+1.22}$  \\
 	&	\ce{^{210}Pb} 	&	$0$	&   \!\!$.1$	&	$(6.48 \pm 0.25) \times 10^{-8}$ 	&	$_{-3.55}^{}$  \\
 	&	\ce{^{235}U} 	&	$0$	&   \!\!$.01$	&	$(5.71 \pm 0.03) \times 10^{-10}$ 	&	$_{-0.31}^{+0.20}$  \\
 	&	\ce{^{210}Pb} 	&	$1$	&   &	$(5.23 \pm 0.19) \times 10^{-8}$ 	&	$_{-0.69}^{+3.15}$  \\
 	&	\ce{^{232}Th} 	&	$10$	&   &	$(1.15 \pm 0.05) \times 10^{-8}$ 	&	$_{-0.64}^{+0.34}$  \\
 	&	\ce{^{238}U} 	&	$10$	&   &	$(8.35 \pm 0.68) \times 10^{-9}$ 	&	$_{-3.96}^{}$  \\
 	&	\ce{^{210}Pb} 	&	$10$	&   &	$(6.85 \pm 0.69) \times 10^{-8}$ 	&	$_{-4.23}^{+4.88}$  \\
 	&	\ce{^{235}U} 	&	$10$	&   &	$(3.84 \pm 0.31) \times 10^{-10}$ 	&	$_{-1.82}^{}$  \\

    {\it \MX} \\ \cline{1-1} \\[-15pt]
 	&	\ce{^{232}Th} 	&	$0$	&   \!\!$.01$	&	$<4.36 \times 10^{-9}$ 	&	  \\
 	&	\ce{^{238}U} 	&	$0$	&   \!\!$.01$	&	$(6.79 \pm 1.32) \times 10^{-8}$ 	&	$_{-6.42}^{}$  \\
 	&	\ce{^{210}Pb} 	&	$0$	&   \!\!$.01$	&	$<2.05 \times 10^{-5}$ 	&	 $<17.11$ \\
 	&	\ce{^{235}U} 	&	$0$	&   \!\!$.01$	&	$(3.12 \pm 0.61) \times 10^{-9}$ 	&	$_{-2.95}^{}$  \\

    {\it \HEX} \\ \cline{1-1} \\[-15pt]
    &	\ce{^{210}Pb} 	&       &	 	        &	$(8.23 \pm 0.20) \times 10^{-4}$ 	&	$_{-6.43}^{+6.43}$  \\
   \end{tabular}
   \label{tab:Results_Surface_contamination}
  \end{table*}
 \end{center}

 \section{Summary}
 \label{sec:summary}
 
 We presented the background model of the CUORE data based on an exposure of 1038.4~kg yr. The data reconstruction is achieved by means of a multiparametric Bayesian fit of 39 spectra encompassing \MI and \MII events across a broad energy range $(0.2, 6.8)$ MeV. Our model describes the observed data well and comparing the results with the CUORE background projections~\cite{Alduino:2017qet} we observe that all components match the expectations except for some surface contamination of the {\it\xtals} and {\it\CuNOSV} volumes. 
 These findings reveal the reliability of the materials screening techniques and emphasize the importance of reliable surface contamination assay and monitoring to mitigate the risk of surface contamination. By subdividing the data in geometric and time slices, we can localize background components to analyze and model their spatial distribution across the detector, and we can study the time evolution of specific contamination. This robust reconstruction of the data over a broad energy range is the basis of forthcoming physics analyses which rely on the continuum spectrum observed in CUORE, for example investigation of the \bbvv spectrum of \ce{^{130}Te} to explore nuclear effects on the spectral shape~\cite{PhysRevLett.123.262501,PhysRevD.100.092002,PhysRevD.107.032006,PhysRevLett.131.162501}.

The information extracted from this background model guides the design and optimization of the CUPID experiment~\cite{CUPID_bsl_paper}, which will exploit the CUORE cryogenic infrastructure to host an array of $^{100}$Mo-enriched scintillating bolometers to search for \bb decay fully exploring the inverted hierarchy of neutrino masses. The scintillating bolometer technique enables event-by-event vetoing of \al-induced events. Therefore, knowledge of the particle origin and locations of background derived from the CUORE background model is crucial to establish the background budget and sensitivity of CUPID. 

 \section*{Acknowledgments} 

The CUORE Collaboration thanks the directors and staff of the Laboratori Nazionali del Gran Sasso
and the technical staff of our laboratories.
This work was supported by the Istituto Nazionale di Fisica Nucleare (INFN);
the National Science Foundation under Grant Nos. NSF-PHY-0605119, NSF-PHY-0500337,
NSF-PHY-0855314, NSF-PHY-0902171, NSF-PHY-0969852, NSF-PHY-1307204, NSF-PHY-1314881, NSF-PHY-1401832, and NSF-PHY-1913374; Yale University, Johns Hopkins University, and University of Pittsburgh.
This material is also based upon work supported by the US Department of Energy (DOE)
Office of Science under Contract Nos. DE-AC02-05CH11231 and DE-AC52-07NA27344;
by the DOE Office of Science, Office of Nuclear Physics under Contract Nos.
DE-FG02-08ER41551, DE-FG03-00ER41138, DE- SC0012654, DE-SC0020423, DE-SC0019316.
This research used resources of the National Energy Research Scientific Computing Center (NERSC).
This work makes use of both the DIANA data analysis and APOLLO data acquisition software packages,
which were developed by the CUORICINO, CUORE, LUCIFER, and CUPID-0 Collaborations.
The authors acknowledge Advanced Research Computing at Virginia Tech for providing computational resources and technical support that have contributed to the results reported within this paper.

\clearpage
\appendix

 \section{\texorpdfstring{\ce{^{40}K} on tower 12}{40K on tower 12}}
 \label{app:K_t12}

\begin{figure}[t]
    \centering
    \includegraphics[width=1.\columnwidth]{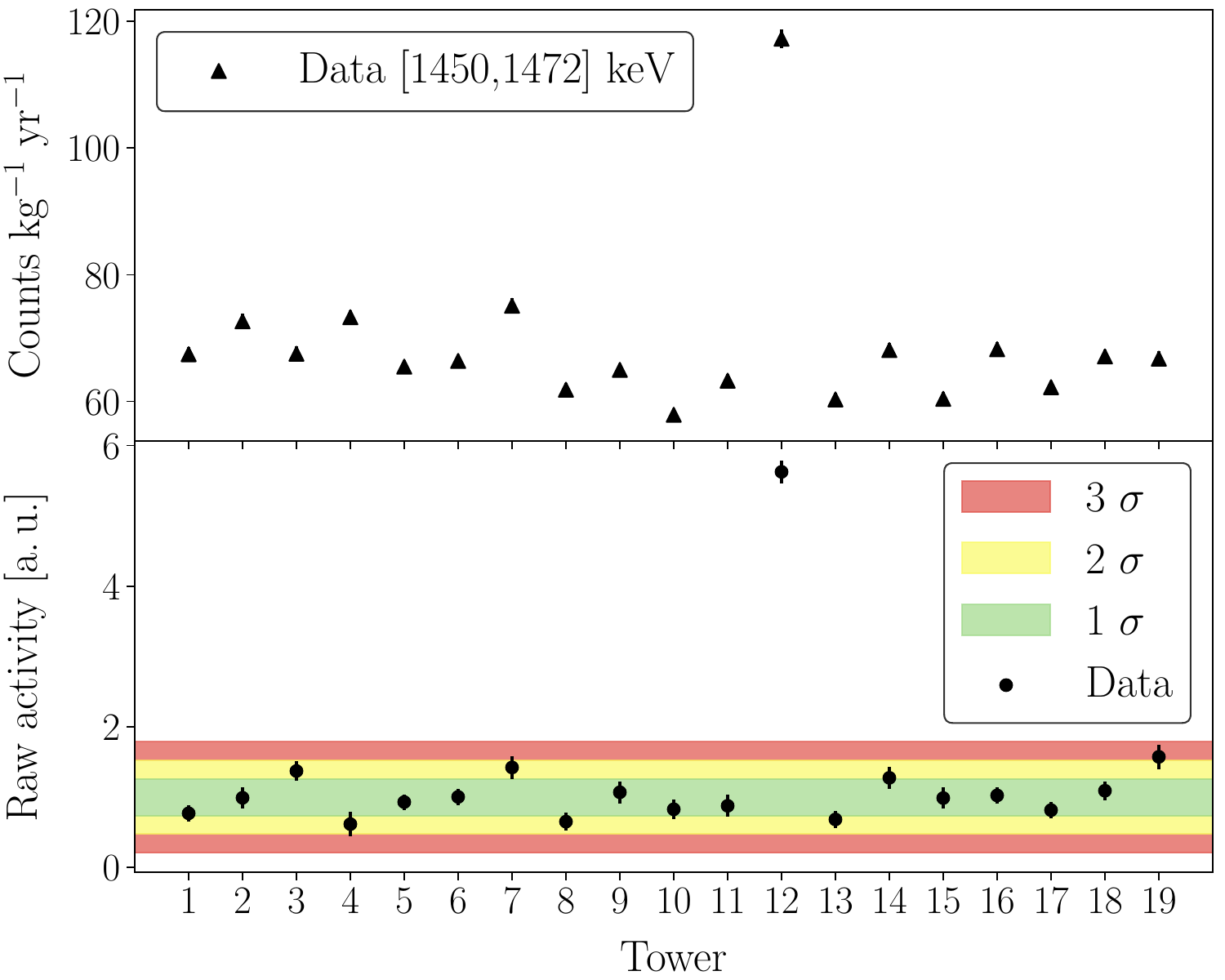}
    \caption{(Top) Yield normalized by the exposure for the potassium 1461-keV line of \ce{^{40}K} for the 19 towers of CUORE. Tower 12 contains approximately double the events compared to the other towers. (Bottom) Normalization factor of the \ce{^{40}K} contamination in the \xtals bulk when performing the fit on the towers individually. The colored bands refer to the 1, 2 and 3 standard deviations coming from the activity distribution when considering all the towers except tower 12.}
    \label{fig:40KT12}
\end{figure}

 We adopted a different treatment for the background contribution of \ce{^{40}K} compared to all the other \MC simulations. Examining the number of counts in the $1461$-keV line of \ce{^{40}K} recorded by each tower, a significant excess is present on tower $12$ (Fig.~\ref{fig:40KT12}).
 To investigate this, we performed the background model fit on each tower separately and extracted the \ce{^{40}K} contribution from each. The result points to a clear surplus coming from the \xtals of tower 12, with an activity around $5.7$ times higher with respect to the average coming from the other towers: no clear deviation from a uniform trend is present for the other volumes.
 Therefore, we decided to include the tower 12 contamination of \ce{^{40}K} with a fixed activity ratio of 5.7 with respect to the remaining towers.
 This procedure allows us to directly include very non-uniform activities directly inside the fit, improving the overall data reconstruction.
 Moreover, since \ce{^{40}K} in {\it\xtals} induces a prominent background in the \gm region and it is correlated with the 2$\nu\beta\beta$, a more precise description of its distribution can in turn lower the systematics induced on the other background components and on the determination of the \ce{^{130}Te} half-life.

\section{Contaminants in \ce{TeO_{2}}}
\label{app:TeO2OtherContaminants}

 When dealing with the {\it\xtals} volume, which is the active component of the detector, we had to meticulously assess whether to incorporate or not specific contaminants associated with \ce{TeO_2}.

 Thanks to the high statistics collected with \CUORE, we observe a peak in the spectrum at \mbox{$\sim 2316$ keV}, which we cannot to associate with any `conventional' \gm-ray emission. The most promising candidate for its origin is \ce{^{147}Sm}, whose contamination in \ce{TeO_{2}} shows a peak-only spectrum at the Q-value of the \al transition at \mbox{$2311$ keV}.
 \ce{^{147}Sm} is a naturally-occurring isotope and it has been previously found as a crystal contaminant in other cryogenic experiments, where it had been possible to perform tagging of \al events~\cite{CUPID:2019lzs}.
 Since a small energy mis-reconstruction (excess) for \al particles in bolometers has always been observed as a result of calibrating the detectors with \gm-ray lines~\cite{Bellini:2010iw}, we usually model this energy surplus via a \QF. Also in this analysis, in the absence of a more satisfactory alternative, we thus simulate a \ce{^{147}Sm} contamination in the bulk of {\it \xtals} and later add an {\it ad hoc} \QF.

 We conducted dedicated investigations on possible fallout products in \ce{TeO_{2}},namely \ce{^{137}Cs} and \ce{^{90}Sr}, which are always produced together.
 \ce{^{137}Cs} has been found in traces in the copper used for \CUORE. This isotope exhibits a characteristic \gm-ray line at $661.7$ keV, which can be used to constrain its activity.
 Including a \ce{^{137}Cs} contamination gives an activity $\lesssim 70$ nBq kg$^{-1}$ making its contribution and effects negligible. Therefore, we decided not to include this contaminant in the {\it\xtals} volume.
 \ce{^{90}Sr} undergoes two consecutive pure \bt decays, first to \ce{^{90}Y} and then to \ce{^{90}Zr}, with Q-values of $546.0$ keV and $2280.1$ keV, respectively. 
 The resulting spectrum is featureless and degenerate to that by the \bbvv decay of \ce{^{130}Te}, hence inducing a strong correlation with it.
 However, due to its short half-life (28.8 yr), its concentration is expected to be smaller than $10^{-20}$ g/g, a sensitivity that cannot be reached by any material screening technique at present.
 Moreover, the goodness of fit with a background model including \ce{^{90}Sr} does not improve. We therefore decide not to include it in the reference fit but, since this study is devoted to the determination of the possible experimental contamination, we study the systematics induced by its addition to be conservative. 

 \section{Identified peaks}
 \label{app:peaks}
 The identification of contaminants in the \CUORE experiment is mainly based on the recognition of the correspondent characteristic spectral features, such as a peak search in the data. Moreover, the same procedure has been employed to define both the diagonal bands used to generate the spectra for the data reconstruction, and the binning for energy histograms.

 Therefore, we identified all the peaks in the \gm region for both \MI and \MII energy spectra as well as the \al peaks in the \MII total-energy data and report the results in Tables~\ref{tab:M1VisibleGammas}, \ref{tab:M2BandsVisibleGammas}, \ref{tab:M2BandsListGamma} and~\ref{tab:M2BandsList}.
 \begin{table}[t]
  \caption{List of identified gamma peaks in the \gm region of \MI spectrum. {\it SE} and {\it DE} refer to single-escape and double-escape peaks, respectively.}
  \label{tab:M1VisibleGammas}
  
  \centering
  \begin{tabular}{r|r}
   E [keV]    &Nuclide        \\[3pt]
   \hline \\[-8pt]
   $ 238.6$\, &\ce{^{212}Pb}  \\
   $ 295.2$\, &\ce{^{214}Pb}  \\
   $ 338.3$\, &\ce{^{228}Ac}  \\
   $ 351.9$\, &\ce{^{214}Pb}  \\
   $ 427.8$\, &\ce{^{125}Sb}  \\
   $ 433.9$\, &\ce{^{108m}Ag} \\
   $ 463.0$\, &\ce{^{228}Ac}  \\
   $ 511.0$\, &\ce{e+e-}      \\
   $ 583.2$\, &\ce{^{208}Tl}  \\
   $ 609.3$\, &\ce{^{214}Bi}  \\
   $ 614.3$\, &\ce{^{108m}Ag} \\
   $ 657.7$\, &\ce{^{110m}Ag} \\
   $ 665.4$\, &\ce{^{214}Bi}  \\
   $ 722.9$\, &\ce{^{108m}Ag} \\
   $ 727.3$\, &\ce{^{212}Bi}  \\
  \end{tabular}
  \quad
  \begin{tabular}{r|r}
   E [keV]    &Nuclide        \\[3pt]
   \hline \\[-8pt]
   $ 768.4$\, &\ce{^{214}Bi}  \\
   $ 794.9$\, &\ce{^{228}Ac}  \\
   $ 803.0$\, &\ce{^{210}Po}  \\
   $ 834.8$\, &\ce{^{ 54}Mn}  \\
   $ 860.6$\, &\ce{^{208}Tl}  \\
   $ 911.2$\, &\ce{^{228}Ac}  \\
   $ 934.1$\, &\ce{^{214}Bi}  \\
   $ 964.0$\, &\ce{^{228}Ac}  \\
   $ 969.0$\, &\ce{^{228}Ac}  \\
   $1001.0$\, &\ce{^{234m}Pa} \\
   $1063.6$\, &\ce{^{207}Bi}  \\
   $1120.3$\, &\ce{^{214}Bi}  \\
   $1173.2$\, &\ce{^{ 60}Co}  \\
   $1238.1$\, &\ce{^{214}Bi}  \\
   $1332.5$\, &\ce{^{ 60}Co}  \\
  \end{tabular}
  \quad
  \begin{tabular}{r|r}
   E [keV]    &Nuclide        \\[3pt]
   \hline \\[-8pt]
   $1238.1$\, &\ce{^{214}Bi}  \\
   $1460.8$\, &\ce{^{ 40} K}  \\
   $1588.2$\, &\ce{^{228}Ac}  \\
   $1620.5$\, &\ce{^{212}Bi}  \\
   $1630.6$\, &\ce{^{228}Ac}  \\
   $1729.6$\, &\ce{^{214}Bi}  \\
   $1764.5$\, &\ce{^{214}Bi}  \\
   $1847.4$\, &\ce{^{214}Bi}  \\
   $2103.5$\, &\ce{^{208}Tl_{\it SE}} \\
   $2118.5$\, &\ce{^{214}Bi}  \\
   $2204.1$\, &\ce{^{214}Bi}  \\
   $2316.5$\, &\ce{^{147}Sm}  \\
   $2447.9$\, &\ce{^{214}Bi}  \\
   $2505.6$\, &\ce{^{ 60}Co}  \\
   $2614.5$\, &\ce{^{208}Tl}  \\
  \end{tabular}
 \end{table}
 
 \begin{table}[t]
  \caption{List of identified gamma peaks for the \MII gamma bands.}
  \label{tab:M2BandsVisibleGammas}
  \centering
  \begin{tabular}{r|r}
   E [keV]    &Nuclide        \\[3pt]
   \hline \\[-8pt]
   $ 328.0$\, &\ce{^{228}Ac}  \\
   $ 351.9$\, &\ce{^{214}Pb}  \\
   $ 409.5$\, &\ce{^{228}Ac}  \\
   $ 427.9$\, &\ce{^{125}Sb}  \\
   $ 434.2$\, &\ce{^{108m}Ag} \\
   $ 511.0$\, &\ce{e+e-}      \\
   $ 583.2$\, &\ce{^{208}Tl}  \\
   $ 609.3$\, &\ce{^{214}Bi}  \\
   $ 722.9$\, &\ce{^{110m}Ag} \\
   $ 768.4$\, &\ce{^{214}Bi}  \\
   $ 794.9$\, &\ce{^{228}Ac}  \\
  \end{tabular}
  \quad
  \begin{tabular}{r|r}
   E [keV]    &Nuclide        \\[3pt]
   \hline \\[-8pt]
   $ 821.5$\, &\ce{^{ 60}Co_{\it SE}} \\
   $ 835.7$\, &\ce{^{228}Ac}  \\
   $ 911.2$\, &\ce{^{228}Ac}  \\
   $ 950.0$\, &\ce{^{ 40} K_{\it SE}} \\
   $ 969.0$\, &\ce{^{228}Ac}  \\
   $1120.3$\, &\ce{^{214}Bi}  \\
   $1173.2$\, &\ce{^{ 60}Co}  \\
   $1332.5$\, &\ce{^{ 60}Co}  \\
   $1592.5$\, &\ce{^{208}Tl_{\it DE}} \\
   $1764.5$\, &\ce{^{214}Bi}  \\
   $2103.5$\, &\ce{^{208}Tl_{\it SE}} \\
  \end{tabular}
 \end{table}

 \begin{table}[t]
  \caption{List of \gm emitters used to define the \MII total-energy bands in the \gm region.}
  \label{tab:M2BandsListGamma}
  \begin{tabular}{r|r}
   E [keV]    &Nuclide        \\[3pt]
   \hline \\[-8pt]
     511.0\,       &e$^{+}$e$^{-}$    \\
     583.2-609.3\,     &\ce{^{208}Tl}-\ce{^{214}Bi}  \\
     722.9-727.3\,     &\ce{^{108m}Ag}-\ce{^{212}Bi} \\
     834.8\,     &\ce{^{54}Mn}   \\
     911.2\,     &\ce{^{228}Ac}  \\
     969.0\,        &\ce{^{228}Ac}  \\
    1063.7\,        &\ce{^{207}Bi}  \\
    1120.3\,        &\ce{^{214}Bi}  \\
    1173.2\,        &\ce{^{60}Co}   \\
    1238.1\,        &\ce{^{214}Bi}  \\
    1332.5\,        &\ce{^{60}Co}   \\
    1377.7-1408.0\,       &\ce{^{214}Bi}  \\
  \end{tabular}
  \hspace{4pt}
  \begin{tabular}{r|r}
   E [keV]    &Nuclide        \\[3pt]
   \hline \\[-8pt]
    1460.5\,        &\ce{^{40}K}    \\
    1509.2\,        &\ce{^{214}Bi}  \\
    1588.2\,        &\ce{^{228}Ac}  \\
    1620.5-1630.6\,       &\ce{^{212}Bi}-\ce{^{228}Ac}  \\
    1661.3\,        &\ce{^{214}Bi}  \\
    1729.6\,        &\ce{^{214}Bi}  \\
    1764.5\,        &\ce{^{214}Bi}  \\
    1847.4\,        &\ce{^{214}Bi}  \\
    2103.5-2118.6\,       &\ce{^{208}Tl_{\it SE}}\,-\ce{^{214}Bi} \\
    2204.1\,        &\ce{^{214}Bi}  \\
    2447.9-2505.6\,       &\ce{^{214}Bi}-\ce{^{60}Co}   \\
    2614.5\,        &\ce{^{208}Tl}  \\
  \end{tabular}
 \end{table}

 \begin{table}[t]
  \caption{List of \al emitters used to define the \MII total-energy bands in the \al region.}
  \label{tab:M2BandsList}
  \centering
  \begin{tabular}{r|r}
   Q-value [keV]    &Nuclide        \\[3pt]
   \hline \\[-8pt]
   $4081.6$\, &\ce{^{232}Th}  \\
   $4269.7$\, &\ce{^{238} U}  \\
   $4770.0$\, &\ce{^{230}Th}  \\
   $4857.7$-$4870.6$\, &\ce{^{234} U}-\ce{^{226}Ra}  \\
   $5407.5$\, &\ce{^{210}Po}  \\
  \end{tabular}
  \quad
  \begin{tabular}{r|r}
   Q-value [keV]    &Nuclide        \\[3pt]
   \hline \\[-8pt]
   $5520.1$\, &\ce{^{228}Th}  \\
   $5590.3$\, &\ce{^{222}Rn}  \\
   $5788.9$\, &\ce{^{224}Ra}  \\
   $6114.7$\, &\ce{^{218}Po}  \\
   $6207.4$\, &\ce{^{212}Bi}  \\
  \end{tabular}
 \end{table}

\bibliography{ref}

\end{document}